# Anisotropic Water Structure at Charged Interfaces Studied by Depth Resolved Vibrational SFG/DFG Spectroscopy

Álvaro Diaz-Duque[1], Vasileios Balos[2], Martin Wolf[1], Alexander P. Fellows[1], and Martin Thämer[1*].

[1] Fritz-Haber-Institut der Max-Planck-Gesellschaft, Faradayweg 4-6, 14195, Berlin, Germany

[2] Instituto Madrileño de Estudios Avanzados en Nanociencia (IMDEA Nanociencia), 28049, Madrid, Spain

* Corresponding author

thaemer@fhi-berlin.mpg.de

(tel.): +49 (0)30 8413 5220

## Abstract

The molecular water structure at charged aqueous interfaces is shaped by interfacial electric fields, which can induce significant anisotropy in the molecular orientations extending over nanometer-scale distances. Despite great relevance, very little is known about the details of this depth-dependent anisotropic water structure, mainly due to the lack of appropriate experimental techniques. Here, we present a depth-resolved study of the water anisotropy at the interface to insoluble charged surfactants using a newly developed technique which allows for directly correlating nonlinear vibrational spectra with depth information on the nanometer scale. We demonstrate that the obtained data allows for a full reconstruction of the nonlinear vibrational responses as function of depth. The results for the case of low salinity solutions show the presence of two pronounced regions within the interfacial anisotropy with largely deviating degrees of preferential molecular orientations. A spectral analysis of the depth-dependent vibrational responses furthermore reveals that the natural local hydrogen-bond structure of bulk water remains largely unperturbed throughout the interfacial region, including water in direct proximity of the surface charges. These findings significantly refine our understanding of the anisotropic water structure at the interface to hydrophilic charged surfactants and showcase the large potential of our depth-resolved spectroscopic technique.

## Introduction

The elucidation of the depth-dependent molecular water structure at charged aqueous interfaces is a central goal in fundamental interface research because of its overwhelming relevance in a broad variety of fields, ranging from biophysics to environmental chemistry and electrochemistry.[1–6] Particularly important for environmental chemistry are interfaces between water and charged surfactants as they represent the majority of the oceans surfaces and can also be found in aqueous aerosols.[7,8] The presence of a net surface charge leads to reorientation of water molecules in the subsurface region and thus influences the interfacial molecular structure.[9–11] This process is induced by the electrostatic field that penetrates the subsurface region and decays towards the bulk. The resulting preferential orientation makes the molecular structure anisotropic, which has important consequences for the thermodynamic properties of the interface as e.g. molecular orientation generally leads to a lowering of entropy.[12,13] Such ordering effects have therefore crucial impact on many aspects of interfacial chemistry such as transport processes across the interface, macroscopic surface properties such as surface tension, as well as modulations of local electrochemical potentials. [14–19]

The thickness of the structurally anisotropic region is thereby highly variable and can reach up to hundreds of nanometers. The length-scale of the anisotropy decay depends, for example, on the salt concentration of the electrolyte which modulates the screening of the electrostatic field. At high ionic strengths, the anisotropy will decay faster as function of depth than for low ionic strengths.[20] Determining the exact evolution of the water structure with depth has, however, proven extremely challenging, both theoretically and experimentally.[21–26] While the evolution of the electrostatic potential with depth can be described by diverse theoretical models such as Gouy-Chapman, Gouy-Chapman-Stern, and further refining modifications,[20,27] these theories do not provide direct information on the resulting anisotropic water structure as they treat water as a homogeneous medium. Furthermore, a detailed experimental elucidation of the interfacial water structure has so far been hampered by the lack of a direct probe of the depth-dependent anisotropy. In consequence, our understanding of these structural details is so far highly limited. Open questions include, for example, whether the water structure in direct contact with the surface charges is fundamentally different than the water structure in the layers below.

An experimental technique that has been widely used to study water structures at charged aqueous interfaces is phase-sensitive vibrational sum frequency generation (SFG) spectroscopy.[22,26,28–34] The benefits of this technique for such studies are twofold: the vibrational water spectra are highly sensitive to the hydrogen bond environment, with strong hydrogen bonds leading to red-shifted OH stretch resonances, and weak hydrogen bonds leading to blue-shifts.[35,36] That way, vibrational water spectra allow for the characterization of the H-bond network. Secondly, due to the special selection rules governing the second-order light-matter interactions, the technique is sensitive to anisotropic molecular arrangements (within the electric dipole approximation).[37–41] In SFG the sign of the response is directly related to the molecular orientation forming peaks or dips in the spectra.[42–44] This property makes the signals from isotropic regions vanish (via cancellation), which allows for exclusively probing the regions possessing structural anisotropy such as the interfacial water at charged interfaces. As the anisotropy in these aqueous systems is dominated by preferential molecular orientation, the second-order susceptibility, $\chi^{(2)}$, which is the quantity of interest in SFG spectroscopy, serves as a direct measure of its extent. A clear limitation of this technique in its traditional form is, however, that the measured response is the integration over the entire anisotropic region, thus the important evolution of the anisotropy with depth , $\chi^{(2)}(z)$, is inaccessible.[45] Acknowledging this deficiency, several advanced experimental approaches have been developed[22,24,26,45–47] which indeed provide further insight into the depth-dependent water structure. However, due to their intrinsically limited depth-sensitivities and/or the requirements for extended theoretical models and often contentious assumptions, these approaches have so far not managed to obtain a conclusive depth resolved picture of water at charged interfaces.

Recently, we have presented a new concept in second-order vibrational spectroscopy that inherently yields a combination of nonlinear vibrational spectra of molecular interfaces with depth information on the sub-nm scale.[45,48–50] The technique is based on the simultaneous, phase-sensitive measurement of sum- and difference-frequency generation signals (SFG and DFG) and enables the decomposition of signal contributions from different depths within the interfacial region. The desired depth information is thereby encoded in the phase and amplitude differences between the resulting SFG and DFG spectra. Because of its capability to provide precise depth information, the use of this experimental approach to characterize the anisotropy at charged aqueous interfaces bears high potential. Nonetheless, such implementation is far from trivial. As the desired depth information must be extracted from small changes in relative phase and amplitude between SFG and DFG signals, extraordinary accuracy and precision of the measurement is required. Achieving this for measurements on aqueous interfaces that are constantly in motion and yield extremely small SFG signals is obviously highly challenging.

In this contribution, we present an SFG/DFG spectrometer that fulfills these requirements, and we demonstrate its suitability for depth-resolved studies of charged aqueous interfaces by extracting and analyzing the vibrational responses of anisotropic water within the first nanometer of the interface and the region below. In the first part, we describe the theory underlying this technique, followed by a presentation of the experimental details of the approach. In the second part, we present depth-resolved measurements of aqueous electrolyte solutions with insoluble charged surfactants. From the data obtained in these measurements, we successfully reconstruct the entire depth-dependent second-order susceptibility quantitatively, which yields crucial insights into the depth-dependent molecular structure in these systems.

## Theory

The concept of depth-resolved nonlinear vibrational spectroscopy is based on the simultaneous measurement of phase-resolved SFG and DFG spectra from a sample of interest. While the basic principle of this concept is described elsewhere,[45,48,49,51] here we present the full theoretical framework of this technique from an experimental perspective (main text and supplementary information) which includes the mathematical and graphical description of the nonlinear light-matter interaction for ultrashort laser pulses:

In vibrational second-order spectroscopy, a nonlinear signal ($\widetilde{\mathrm{E}}_{(\omega_\rho)}$) is generated by the frequency-mixing of two input fields $\widetilde{\mathrm{E}}_{(\omega_{\mathrm{IR}})}$ and $\widetilde{\mathrm{E}}_{(\omega_{\mathrm{vis}})}$. The response follows the general relation:[50,52]

Equation 1.

$$\widetilde{\mathrm{E}}_{(\omega_\rho)} \propto \int_{-\infty}^{\infty} \mathrm{d}\omega_{\mathrm{IR}} \int_{-\infty}^{\infty} \mathrm{d}\omega_{\mathrm{vis}} \widetilde{\mathrm{E}}_{(\omega_{\mathrm{IR}})} \widetilde{\mathrm{E}}_{(\omega_{\mathrm{vis}})} \chi^{(2)}_{eff,(\omega_\rho=\omega_{IR}+\omega_{vis})} \cdot \delta(\omega_\rho - \omega_{\mathrm{IR}} - \omega_{\mathrm{vis}})$$

where $\chi^{(2)}_{eff,(\omega_\rho=\omega_{IR}+\omega_{vis})}$ is the effective second-order susceptibility. Because all fields are real in their time domain representation, their frequency axes span from negative to positive infinity. As $\chi^{(2)}_{eff,(\omega_\rho=\omega_{IR}+\omega_{vis})}$ depends on two independent frequencies, it can be expressed in the form of a two-dimensional frequency plane. The overall integral in equation 1 can consequently be split into the following four contributions corresponding to the four quadrants in this 2D frequency plane (figure 1).

Equation 2.

$$\widetilde{\mathrm{E}}_{(\omega_{\mathrm{SFG}})} \propto \int_{0}^{\infty} \mathrm{d}\omega_{\mathrm{IR}} \int_{0}^{\infty} \mathrm{d}\omega_{\mathrm{vis}} \widetilde{\mathrm{E}}_{(\omega_{\mathrm{IR}})} \widetilde{\mathrm{E}}_{(\omega_{\mathrm{vis}})} \chi^{(2)}_{eff,(\omega_\rho=\omega_{IR}+\omega_{vis})} \cdot \delta(\omega_\rho - \omega_{\mathrm{IR}} - \omega_{\mathrm{vis}})$$

Equation 3.

$$\widetilde{E}_{(-\omega_{SFG})} \propto \int_{-\infty}^{0} d\omega_{IR} \int_{-\infty}^{0} d\omega_{vis} \widetilde{E}_{(\omega_{IR})} \widetilde{E}_{(\omega_{vis})} \chi^{(2)}_{eff,(\omega_\rho=\omega_{IR}+\omega_{vis})} \cdot \delta(\omega_\rho - \omega_{IR} - \omega_{vis})$$

Equation 4.

$$\widetilde{E}_{(-\omega_{DFG})} \propto \int_{0}^{\infty} d\omega_{IR} \int_{-\infty}^{0} d\omega_{vis} \widetilde{E}_{(\omega_{IR})} \widetilde{E}_{(\omega_{vis})} \chi^{(2)}_{eff,(\omega_\rho=\omega_{IR}+\omega_{vis})} \cdot \delta(\omega_\rho - \omega_{IR} - \omega_{vis})$$

Equation 5.

$$\widetilde{E}_{(\omega_{DFG})} \propto \int_{-\infty}^{0} d\omega_{IR} \int_{0}^{\infty} d\omega_{vis} \widetilde{E}_{(\omega_{IR})} \widetilde{E}_{(\omega_{vis})} \chi^{(2)}_{eff,(\omega_\rho=\omega_{IR}+\omega_{vis})} \cdot \delta(\omega_\rho - \omega_{IR} - \omega_{vis})$$

where the first two correspond to SFG responses, and the other two to the DFG responses (as indicated in figure 1a), representing complex conjugate pairs. Importantly, the measured SFG and DFG responses are not generally equal, with their relation depending on the resonance conditions and spatial origin of the nonlinear signals, as discussed later. Following the two-dimensional representation of the second-order interaction, the mathematical description of the nonlinear response can be visualized graphically. The nonlinear interaction of the material with the IR beam (included in the figure 1b) corresponds to a multiplication of each row of the susceptibility matrix with the complex spectrum of the IR pulses (vertical interaction). Because of the relation $\omega_\rho = \omega_{IR} + \omega_{vis}$, the nonlinear interaction with the visible beam appears along the diagonal (linearly shifted by the visible center frequency). The multiplication of the matrix with the complex spectrum of the visible beam along this diagonal now generates a tilted region in the matrix that contributes to the nonlinear response. The latter is obtained by simple projection of this region onto the vertical ($\omega_\rho$) axis. This graphical representation is general and fully describes the operations in equation 1. It even includes cases where electronic resonances and vibronic coupling are present as illustrated in figure 1b. The clear benefit of this graphical representation is that it yields a straightforward way of visualizing the nonlinear interactions which lead to the generation of the second order signals e.g. for comparisons of the different experimental approaches to obtain the desired nonlinear vibrational information as shown in the supplementary information.

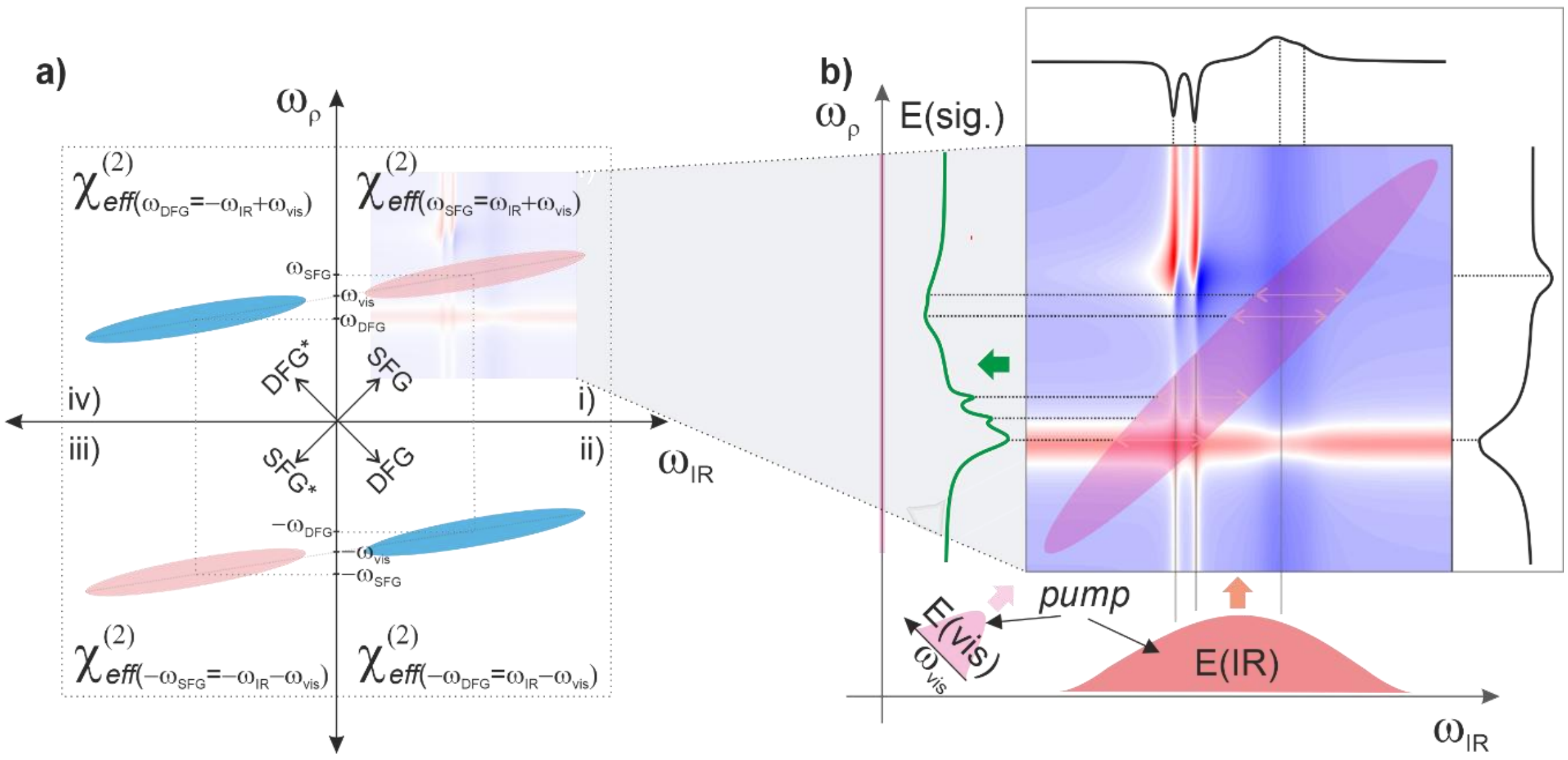

Figure 1. a) Two-dimensional representation of the effective second-order susceptibility as a function of infrared and signal frequencies. SFG signals (red shaded area) and DFG signals (blue shaded area). b) Graphical

representation of second-order nonlinear interactions. Nonlinear signal E(sig.) is obtained by multiplication of the infrared pump field E(IR) and the visible field E(vis) with the nonlinear susceptibility. Plots on the top and right side are projections of the susceptibility indicating different resonances (vibrational resonances along the horizontal and electronic along the vertical axis). For better clarity only the imaginary parts of the susceptibility are shown in the figure.

Besides the resonant information, the two-dimensional effective susceptibility represented in figure 1a also contains the desired depth information. The origin of this lies in the connection between the measured effective susceptibility and the desired depth-dependent local susceptibility ($\chi^{(2)}(z)$, a 3-dimensional quantity) which is given by the following equation

Equation 6.

$$\chi^{(2)}_{eff,(\omega_\rho=\omega_{IR}+\omega_{vis})} = \int_0^\infty dz\, L_{(\omega_p)} L_{(\omega_{vis})} L_{(\omega_{IR})} \chi^{(2)}(z) e^{i\Delta k_z z}$$

where $L_{(\omega_i)}$are the nonlinear Fresnel factors and $\Delta k_z$ is the z-component of the wavevector mismatch ($\Delta k_z^{SFG} = |k_z(\omega_{IR})| + |k_z(\omega_{vis})| + |k_z(\omega_\rho)|$ and $\Delta k_z^{DFG} = |k_z(\omega_{IR})| - |k_z(\omega_{vis})| - |k_z(\omega_\rho)|$ for positive $\omega_{IR}$ in the commonly applied reflection geometry). In typical applications of SFG or DFG, the frequencies are chosen such that the IR beam is resonant with some specific vibration of the sample while the visible frequencies are fully non-resonant.[40,53] In such a case, the intrinsic second-order susceptibility ($\chi^{(2)}(z)$) becomes largely independent on the visible and output frequencies[50] and can be in good approximation replaced by $\chi^{(2)}_{(\omega_{IR})}(z)$ that only depends on IR frequencies. In fact, it is important to note that the quantity of interest in a vibrational SFG experiment (the "SFG spectrum") is the spectrum along this vibrational axis of the second-order susceptibility, and not the spectrum of the generated SFG light. Applying the non-resonant condition at visible frequencies to $\chi^{(2)}(z)$, we obtain two-dimensional frequency planes with the IR resonances along the horizontal dimension while all the values remain constant along the vertical axis (see figure S-1, supplementary information). As a result, also the SFG and DFG responses in each $\chi^{(2)}_{(\omega_{IR})}(z)$ plane that correspond to the same side of the vibrational axis become equal.[50] However, such equality is still not generally given for the measured responses (effective susceptibility, $\chi^{(2)}_{eff,(\omega_\rho=\omega_{IR}+\omega_{vis})}$). As shown in equation 6, the $\chi^{(2)}(z)$ is modulated by the phase factor $e^{i\Delta k_z z}$, which introduces depth-dependent phase shifts to the generated signals. These phase shifts are different for SFG and DFG in magnitude and direction. This discrepancy originates from the different values of $\Delta k_z$ (different sign and magnitude) for SFG and DFG signals as individual k values invert their sign for regions with negative frequency in the two-dimensional representation of $\chi^{(2)}_{eff,(\omega_\rho=\omega_{IR}+\omega_{vis})}$. This makes the effective SFG and DFG responses sensitive to the spatial origin (depth) of the signal, with significant deviations between SFG and DFG spectra for signals from larger depths and overlapping spectra for pure surface signals. Particularly sensitive to depth is thereby the phase difference between the SFG and DFG spectra with ca. 2 deg. per nm.[48] This property can be exploited to extract the desired depth information, as shown in the result section.

**Experimental**

For a successful depth-resolved study, the phase-resolved SFG and DFG spectra need to be obtained at very high accuracy and precision, which places significant requirements for the experimental implementation of such a spectroscopy. The two signals (SFG and DFG) are generated from the same nonlinear process, meaning that every SFG response is always accompanied by a corresponding DFG signal.[39,45,52,54] In consequence, each SFG experiment can in principle be upgraded to a SFG/DFG spectrometer by simply detecting the simultaneously generated DFG signal. Phase-resolution for each of the two responses can then be achieved interferometrically by superimposing the nonlinear signals with corresponding reference beams, so-called local oscillators (LO).[55,56] In consequence, any of the common experimental SFG methods such as narrow- and broadband frequency-domain approaches[57,58] as well as the recently introduced fully broadband time-domain method[55,56] can in principle be used. However, these techniques are not equally suitable for high precision phase-resolved SFG/DFG experiments on aqueous interfaces. As detailed in the supplementary information, only the broadband time-domain method presented here allows for combining the highly beneficial collinear beam geometry with an efficient suppression of parasitic signal contributions (which is a fundamental requirement for such measurements). The concept of this time-domain approach for phase-resolved SFG spectroscopy was published earlier,[48] here we describe the experimental details of the extension of this concept to an SFG/DFG spectrometer.

Mid-infrared and visible (690 nm) pulses are generated from a commercial 1 kHz Ti:Sa laser system (800 nm; <35 fs pulses) including two independent optical parametric amplifiers (see supplementary information). Inside the spectrometer (see figure 2), a small portion of the IR pulse is split out from the main beam path using a KBr beamsplitter. This weak portion is used for the generation of the local oscillators and passes through a pair of free-standing wire grid polarizers (GP-1 and -2) for power control. Subsequently, this beam portion is collinearly overlapped with the visible on a customized incoupling optic (ICO-1). These two pulses are then focused on a thin (20µm) z-cut quartz crystal where the SFG and DFG LOs are generated (for details see supplementary information), which then travel collinearly with the visible pulse. After passing a delay stage, these pulses are overlapped collinearly (ICO-2) with the main portion of the IR and leave the interferometer.

The single beam output of the interferometer contains four pulses, namely the IR and visible pulses, as well as SFG and DFG LOs, and is guided to the sample area. Before reaching the sample, the beam is split into two portions using a 500 Hz oscillating galvo mirror (OM). The two resulting beams are independently focused on the sample and a reference with off-axis parabolic mirrors (OPMs). Using this approach, every other laser pulse probes sample and reference, which allows for efficient suppression of any phase drifts over the course of the experiment.[48] Just before reaching the sample and reference interfaces, the beams pass through LiF windows for the removal of parasitic signal contributions (see supplementary information). The generated SFG/DFG signals, as well as the respective LOs, which are linearly reflected at the sample surface, are collimated by a second set of OPMs and the two beam-paths are recombined on the same oscillating galvo mirror and sent to detection. A dichroic beam splitter separates SFG and DFG frequencies and sets of edge-pass filters isolate the two heterodyned nonlinear signals. At this point, LOs and nonlinear signals are orthogonally polarized.[56] A pair of calcite crystals is used for tuning the temporal overlap of the nonlinear signals with their respective LOs (details are given in the supplementary information). After a combination of waveplates (WPs) and polarizing beam splitters (PBSs), the interference cross-terms between LOs and nonlinear signals are recorded in a balanced detection scheme,[56] for both the SFG and DFG responses.

During a measurement, the delay between the IR and the combined visible and LOs pulses is scanned by moving the delay stage at a constant speed, while recording the resulting intensity on the detectors for each laser pulse. Scanning the delay stage modulates phase and amplitude of SFG and DFG signals with respect to their LOs, yielding separate interferograms. Due to the alternating probing of reference and sample, we obtain four interferograms for each scan, namely SFG and DFG from both the sample and reference. The final phase-resolved spectra are obtained by Fourier transformation of the sample responses and normalization to the reference responses.

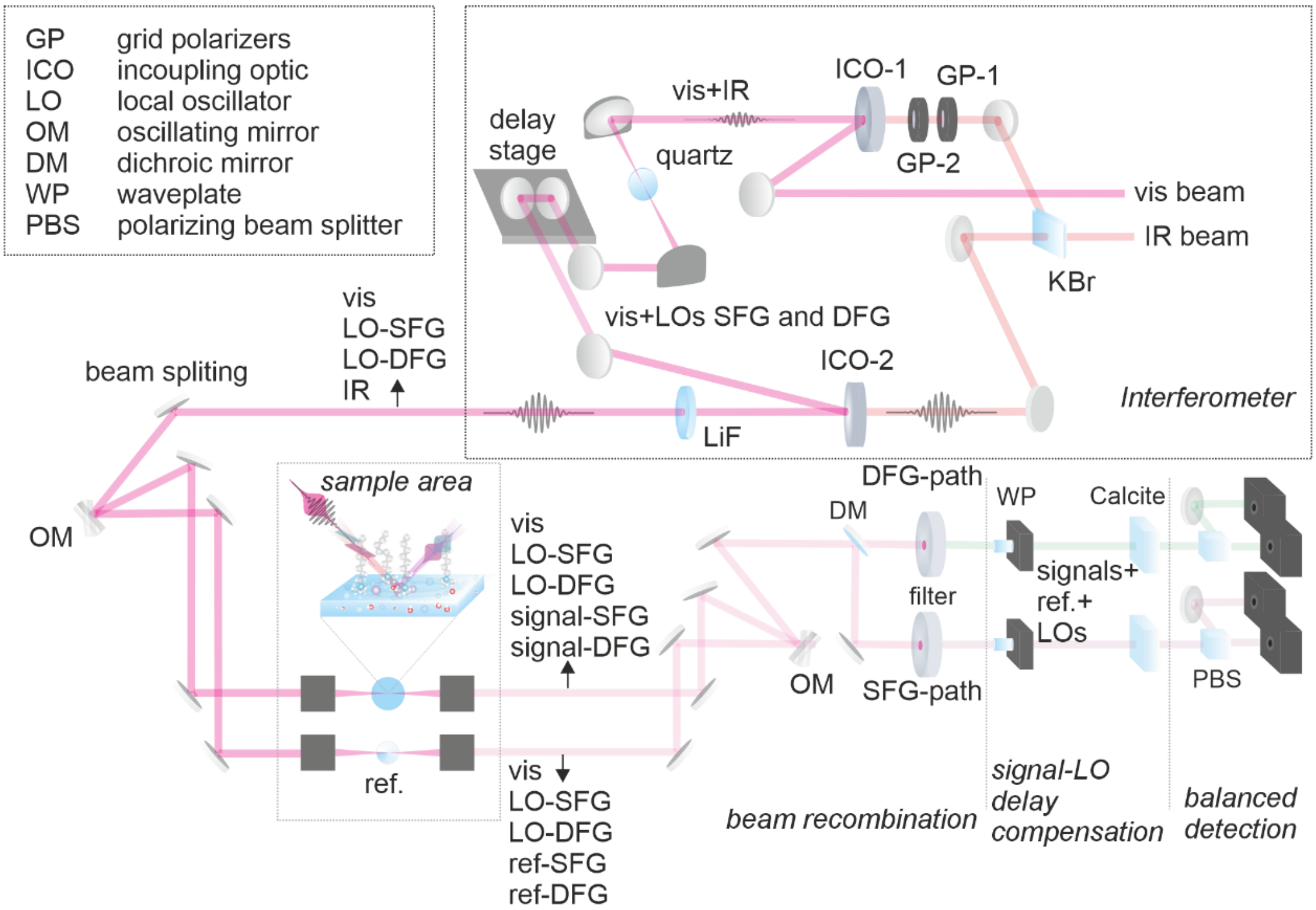

Figure 2. Schematic of the fully collinear phase-resolved SFG and DFG spectrometer

## Results

With this spectroscopic setup in hand, we turn to depth-resolved measurements of interfacial water at the boundary with insoluble charged surfactants. The surfactants employed are dihexadecyldimethylammonium bromide (DHAB) and dihexadecyl phosphate (DHP), which present positively and negatively charged headgroups, respectively, and form densely packed monolayers at the aqueous interface.[59,60] The polarity of the surface charges dictates the water orientation with the molecular dipoles preferentially pointing up for negative and pointing down for positive charges. The resulting anisotropic orientational distribution of water molecules should therefore give rise to SFG/DFG responses with positive (negative) imaginary parts for a net dipole orientation pointing up (down).

Figure 3 shows the imaginary parts of the obtained SFG (red traces) and DFG (blue traces) spectra for aqueous interfaces with DHP (negatively charged, left-side) and DHAB (positively charged, right side), measured in SSP polarization. All the spectra exhibit, at their lower frequencies (below 3000 $cm^{-1}$), pronounced resonant features that correspond to the CH stretching modes from the terminal methyl groups of the surfactant tails. At higher frequencies, there is a broad absorption band between 3000 and 3600 $cm^{-1}$, which corresponds to the OH stretch vibration of the interfacial water molecules with an anisotropic orientational distribution.[61] As expected, a clear sign flip of the OH stretching response can be observed in the spectra when replacing the negative charges with positive charges, confirming the orientational flip of the water molecules for the two different charge polarities.[62] In contrast, for both negatively and positively charged surfactants, the sign of the CH stretching bands remains unchanged. This behavior is expected, since the terminal methyl groups point in the same direction in both systems.

The SFG/DFG measurements of these charged aqueous interfaces are performed at two different salt concentrations in the aqueous solution, with the two samples in the upper panel at $10^{-1}$ M (NaCl), and the solutions in the lower panel at $10^{-5}$ M (NaCl). In the case of high concentration, the electric field from the charges is largely screened and decays on a length scale of approximately 1 nm (Debye length). In consequence, the spatial extent of the water anisotropy should be confined within this thin region. The expected small thickness of the anisotropic layer is clearly confirmed by the significant spectral overlap of SFG and DFG spectra (no considerable phase or amplitude difference between the spectra). In contrast, the field screening in the low concentration case is much weaker with the result that the field reaches much further into the bulk and decays on a length scale of ca. 96 nm.[27] The largely increased thickness of the water layer with preferential molecular orientation in these samples now gives rise to the strong deviations between the SFG and DFG line shapes as shown in the lower panels of figure 3.

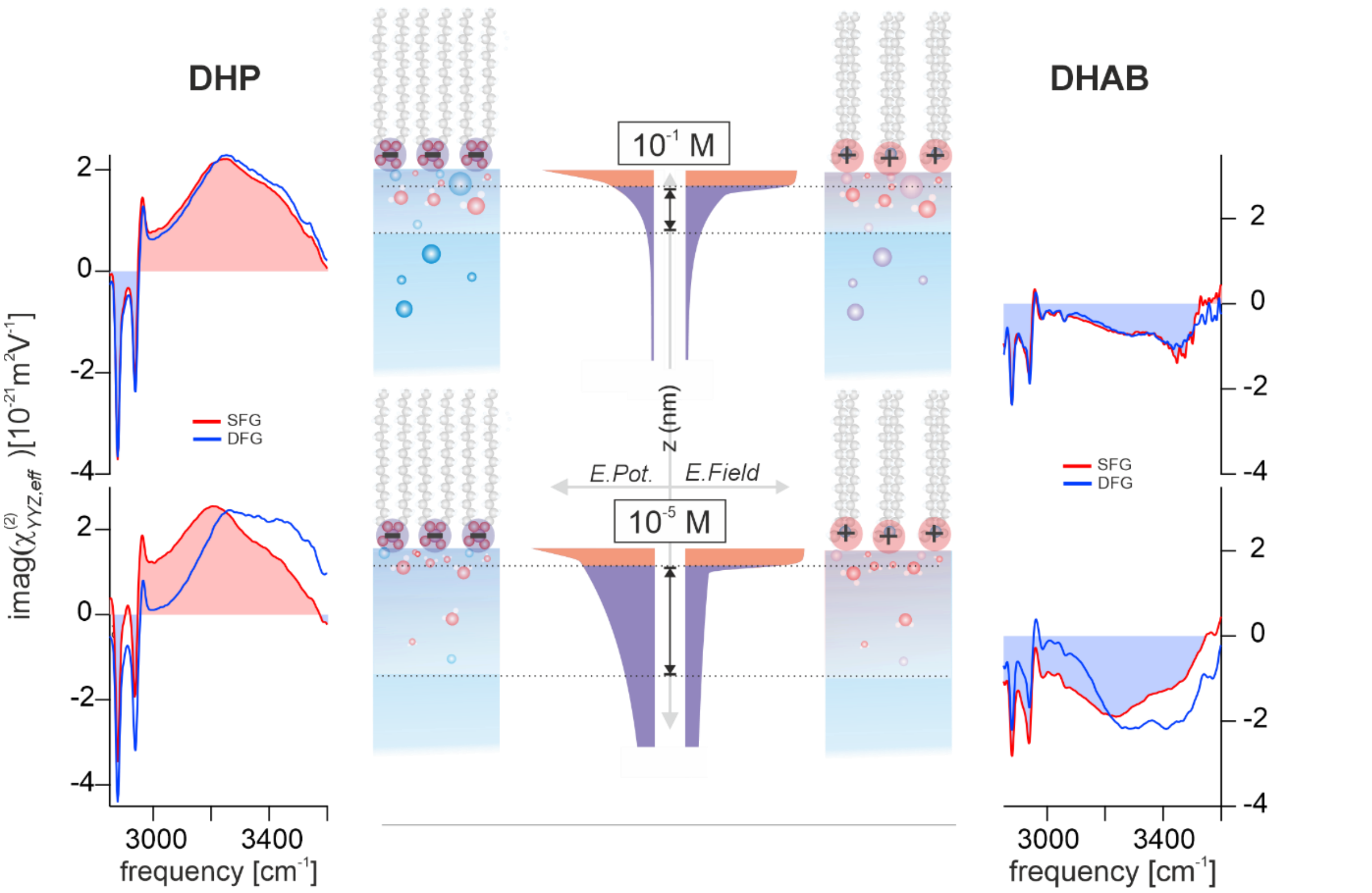

Figure 3. Imaginary SFG (red) and DFG (blue) spectra of charged aqueous interfaces with DHP (dihexadecyl

phosphate, negatively charged, left) and DHAB (dihexadecyldimethylammonium bromide, positively charged, right) at $10^{-1}$ M (upper) and $10^{-5}$ M (lower) NaCl concentrations. Middle panel: schematics of the evolution of the electric potential and field as predicted by the GCS model shown for both salt concentrations. Lower amplitudes in the spectra of the DHAB samples (CH and OH region) indicate a lower surface coverage of DHAB as compared to DHP. Small deviations between SFG and DFG spectra of the $10^{-1}$ M solutions originate from magnetic dipole contributions as detailed in the supplementary information.

The presented measurements clearly confirm the expected behavior of the depth-dependent structural anisotropy under different salinities and demonstrate the sensitivity of the SFG/DFG spectroscopy to these changes in the spatial extent of the anisotropic region. However, for a detailed analysis of the evolution of the structural anisotropy of interfacial water with depth, the full depth-dependent susceptibility $\chi^{(2)}_{(\omega_{IR})}(z)$ needs to be determined from these measurements. In order to do so, we turn to the experiment with negatively charged (phosphate) surfactants at low salt concentrations (bottom left in figure 3) and use a description of the depth-dependent electric potential adopted from the Gouy-Chapman-Stern theory (GCS).[27] The GCS model predicts two electrostatic regimes: a very thin layer directly at the phase boundary, where the electric potential largely drops over the first few water layers (bonded interfacial layer, $z_{BIL}$), followed by an extended region where the potential decays exponentially (diffuse layer, DL). The decay length of the potential in the diffuse layer is thereby given by the Debye length $z_{DL}$. The generated electrostatic field is then given by the negative derivative of the potential and will be relatively large within the BIL followed by a much smaller and exponentially decaying field in the DL (with the same decay constant as for the potential). The amount of preferential molecular orientation of water will somehow follow the evolution of the field and we can consequently also expect $\chi^{(2)}_{(\omega_{IR})}(z)$ to consist of two distinct regimes, $\chi^{(2)}_{BIL}(z)$ and $\chi^{(2)}_{DL}(z)$. Within the DL, water is exposed to a relatively weak electrostatic field, and we can therefore safely assume that the local hydrogen bond structure of water in this region is not altered. In this weak-field region, the $\chi^{(2)}_{DL}(z)$ should be proportional to the electrostatic field and decay exponentially with the Debye length $z_{DL}$.

The exact relation between the electrostatic field and $\chi^{(2)}_{(\omega_{IR})}(z)$ within the BIL is much less clear due to the much higher field strength and potential modifications of the hydrogen bond structure by the solvation of the surfactants headgroups. However, because of the small thickness of the water layer in the BIL ($z_{BIL}$~1 nm), we can describe the evolution of the susceptibility with depth in the BIL by an effective response represented by $\chi^{(2)}_{BIL}$. The overall depth-dependent susceptibility can therefore be described by the following equation.

Eq. 7.

$$\chi^{(2)}_{(\omega_{IR})}(z) = \begin{cases} \chi^{(2)}_{BIL} & 0 < z < z_{BIL} \\ \chi^{(2)}_{DL} \cdot e^{-(z-z_{BIL})/z_{DL}} & z \geq z_{BIL} \end{cases}$$

At this point it is important to note that this notation does not include any assumptions on the depth-dependent evolution of the susceptibility within the BIL. The quantity $\chi^{(2)}_{BIL}$ in the equation above represents the average response of this ca. 1 nm thick region. Furthermore, the DL contribution in equation 7 is often described in the literature by a DC field-induced $\chi^{(3)}$ response, i.e. $\chi^{(3)}_{DL} E_{DC}$.[63–65] The description used here is in principle equivalent with this more common notation but expresses the response in terms of the anisotropic structure resulting from the field-induced reorientation, i.e. the second-order response $\chi_{DL}^{(2)}$. We believe that this choice better highlights the fact that SFG directly probes the structural anisotropy of water rather than the DC field.

Inserting equation 7 into equation 6 yields following results for the predicted SFG and DFG spectra (exploiting the relation $z_{BIL} \ll \Delta k_z$):

Eq. 8.

$$\chi_{SFG,eff}^{(2)} = \chi_{BIL}^{(2)} z_{BIL} + \chi_{DL}^{(2)} z_{DL} \cdot C_{SFG}(z_{DL}, \Delta k_z^{SFG}) \cdot e^{i \cdot atan(\Delta k_z^{SFG} \cdot z_{DL})}$$

Eq. 9.

$$\chi_{DFG,eff}^{(2)} = \chi_{BIL}^{(2)} z_{BIL} + \chi_{DL}^{(2)} z_{DL} \cdot C_{DFG}(z_{DL}, \Delta k_z^{DFG}) \cdot e^{i \cdot atan(\Delta k_z^{DFG} \cdot z_{DL})}$$

with $C(z_{DL}, \Delta k_z) = 1/\sqrt{1 + {z_{DL}}^2 {\Delta k_z}^2}$. Note, the nonlinear Fresnel factors from equation 6 are here absorbed into the effective susceptibilities (see supplementary information for details). From the equations it becomes clear that the BIL contributes equally to SFG and DFG responses, while the DL contribution deviates in phase and amplitude due to the opposite sign and different magnitude of $\Delta k_z$.

In consequence, the BIL contribution cancels in the difference between the measured SFG and DFG spectra, which allows for the isolation of the DL contribution. The resulting residual complex difference spectrum is given by

Eq. 10.

$$\Delta\chi_{eff}^{(2)} = \chi_{DL}^{(2)} z_{DL} \left( C_{SFG}(z_{DL}, \Delta k_z^{SFG}) \cdot e^{i \cdot atan(\Delta k_z^{SFG} \cdot z_{DL})} - C_{DFG}(z_{DL}, \Delta k_z^{DFG}) \cdot e^{i \cdot atan(\Delta k_z^{DFG} \cdot z_{DL})} \right)$$

The term in brackets is unitless and depends on two key parameters: the wavevector mismatch $\Delta k_z$ and the Debye length $z_{DL}$, both of which can be accurately calculated. For the given experimental geometry we obtain for $\Delta k_z^{SFG} = 0.021 nm^{-1}$ and for $\Delta k_z^{DFG} = -0.013 nm^{-1}$ and for a NaCl concentration of $10^{-5}$M the Debye length is $z_{DL} = 96$ nm. A simple division of the difference spectrum by the obtained complex value for this term yields the integrated DL spectrum. From this result, the effective DL contribution to the SFG (DFG) spectrum can be calculated using equation 8 (9), which can then be subtracted from the measured SFG (DFG) spectrum to yield the BIL contribution. With this simple procedure, both the DL and the BIL spectra can be isolated and, using equation 7, the entire evolution of the second-order susceptibility with depth can be reconstructed (details are shown in the supplementary information).

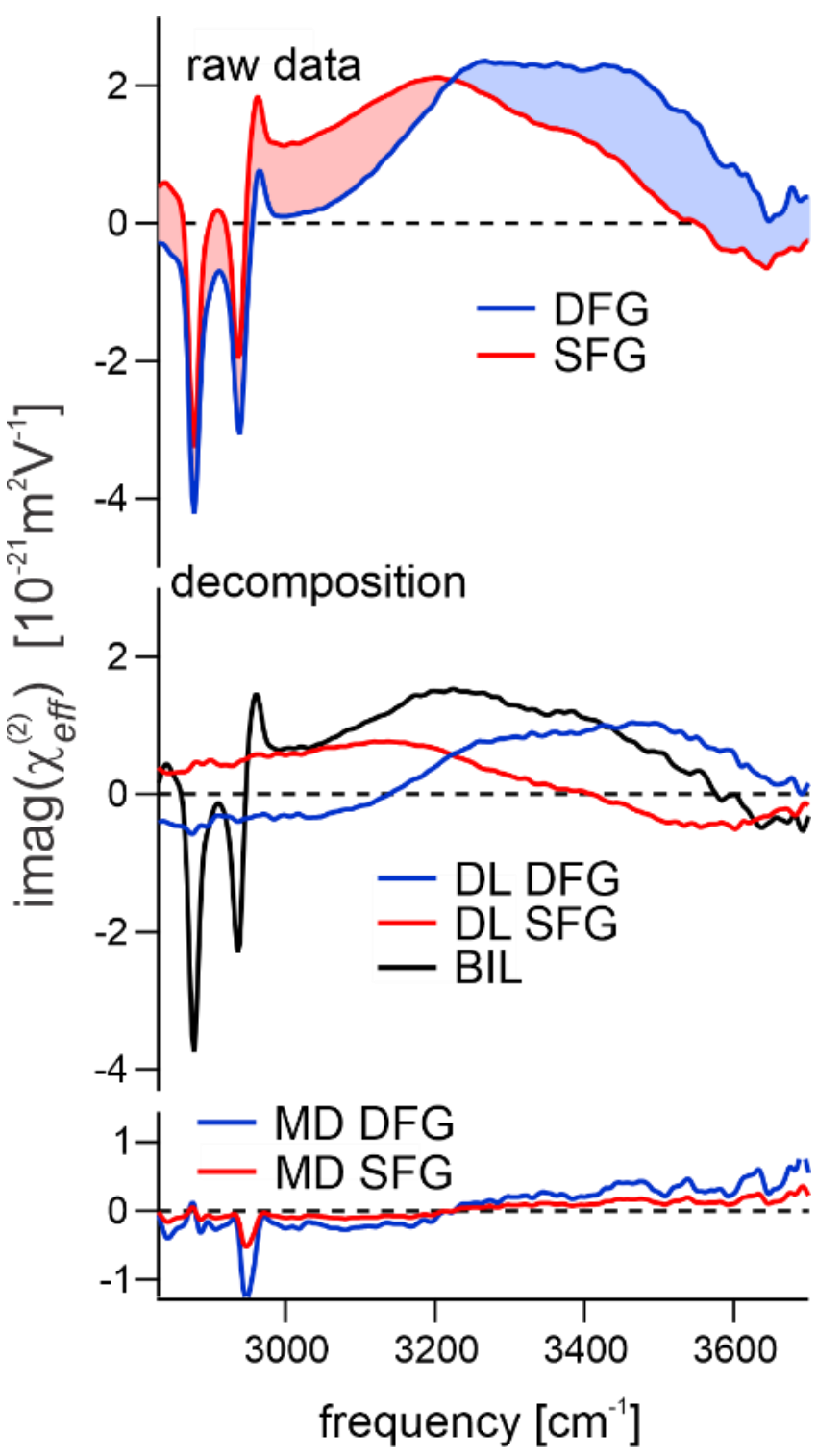

Figure 4. Decomposition of the measured SFG (red) and DFG (blue) spectra (DHP, c(NaCl) = $10^{-5}$M) into the different effective contributions. Upper panel shows raw data, middle panel shows the contribution from the BIL (black, equal for SFG and DFG), and the corresponding DL contributions. The lower panel shows residual SFG and DFG contributions that originate from magnetic dipole transitions (see supplementary information).

Figure 4 shows the decomposition of the measured spectra into the different contributions, namely BIL, DL, and the small residual component that originates from magnetic dipole transitions, MD.[66] For the purpose of the presented analysis the MD signals represent a parasitic contribution which is consequently removed from the BIL and DL spectra (a more detailed description of the MD contributions and their experimental isolation is given in the supplementary information). The decomposition shows that the measured spectra consist of a common contribution from the BIL (black trace lower panel), while the contributions from the DL are clearly phase-shifted and differ for SFG and DFG due to the opposite direction of the phase rotation. The contributions from the MD are meanwhile small and mainly contribute to the CH resonances.

In order to compare DL and BIL spectra, the effect of the depth-related phase-shifts in the DL spectra must be removed, which can be done using equations 8 and 9. The resulting spectra are shown in figure 5a. The BIL spectrum (purple) represents the nonlinear vibrational response from the first approx. 1 nm of the sample ($\int_0^{z_{BIL}} \chi^{(2)}_{BIL}(z)$), while the DL spectrum (black) shows the nonlinear vibrational response beyond this near-surface region ($\int_{z_{BIL}}^{\infty} \chi^{(2)}_{DL}(z)$). A clear discrepancy between the two spectra is the fact that the resonant features from the surfactants ($CH_3$ resonances) only appear in the BIL spectra. This intuitively follows expectations since the surfactant is highly insoluble in water and therefore only decorates the water surface. Beyond this, both spectra show a broad resonance peak centered at around 3200 $cm^{-1}$ which corresponds to the O-H stretch vibration of anisotropic water. Comparison of these two water spectra shows surprising similarities in line-shape. Because of the strong dependency of the O-H stretching frequency on H-bond strength, this spectral similarity is clear evidence that, despite the proximity of these water molecules to the charged head groups, the H-bond environment is not significantly altered. On the contrary, it appears to be comparable to the local H-bond structure in the DL and thus similar to bulk water. In other words, there is no clear indication of any effect of the surface charges on the local H-bond structure of the water layers just below the surface.

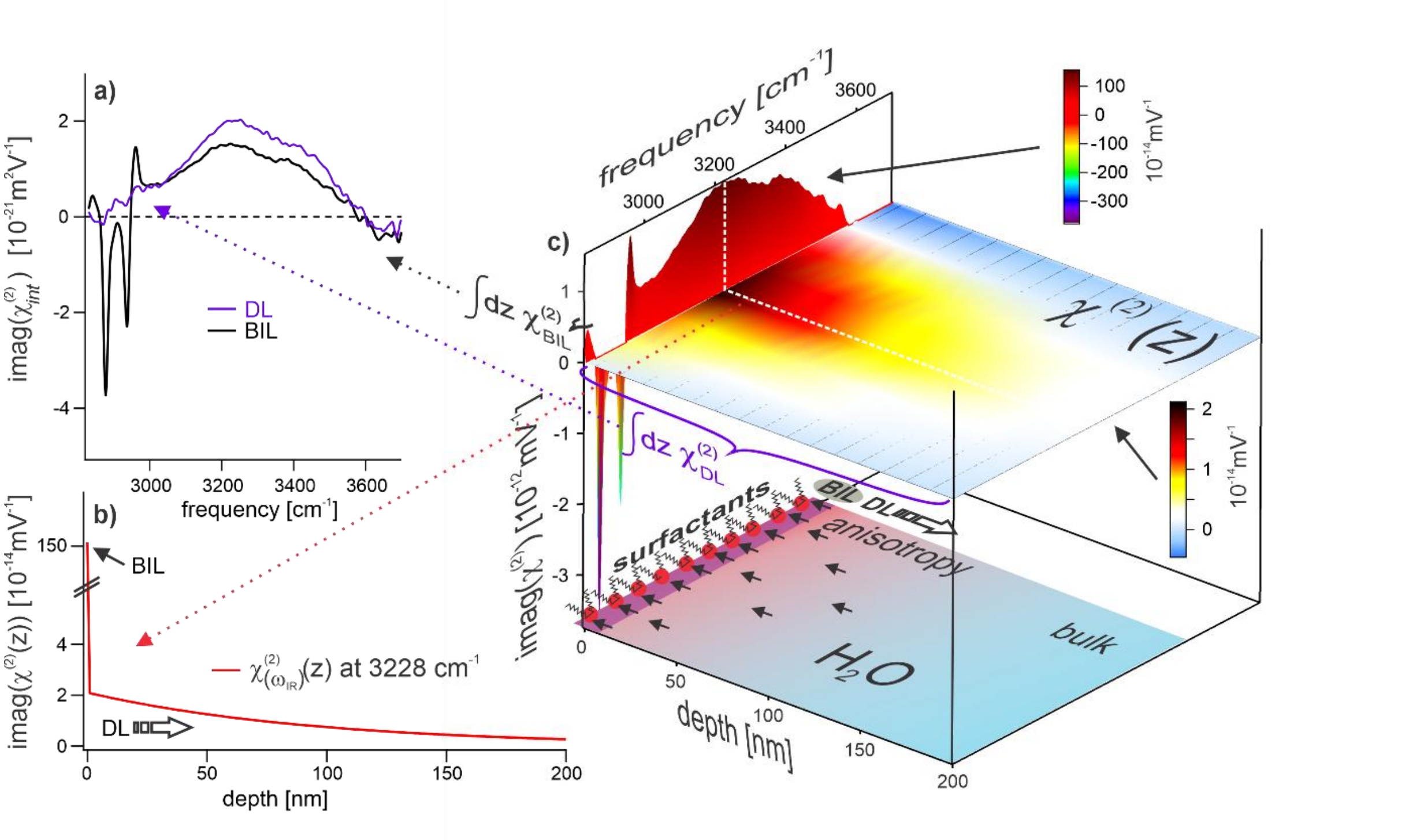

Figure 5. Depth-resolved second-order susceptibility at the charged water interface (DHP, c(NaCl) = $10^{-5}$ M). (a) imaginary parts of the depth-integrated susceptibilities for BIL (black) and DL (purple), respectively. (b): evolution of the susceptibility with the z evaluated at the peak maximum of the water band (3228 cm$^{-1}$). (c) full 3D representation of the depth-dependent susceptibility (imaginary part). Note, the spectral contribution inside the thin BIL represents the average spectrum within this layer and consequently does not show any z-dependence.

From this qualitative analysis of the water responses from the two interfacial regions we now turn to the quantitative analysis of the water anisotropy. Figure 5c shows the resulting full depth-dependent susceptibility $\chi^{(2)}_{(\omega_{IR})}(z)$ which gives direct access to the evolution of the structural anisotropy of water with depth. This 3D plot reveals that the majority of the water anisotropy is clearly located in the BIL, where relatively large values for the susceptibility are obtained compared to much smaller amplitudes in the DL. The reason why the contributions from BIL and DL nevertheless yield similar depth-integrated susceptibilities (as shown in Figure 5a), originates from the very different thicknesses of water that constitute the two regions (~1 nm for the BIL vs. 96 nm in the DL). The large discrepancy in amplitude of $\chi^{(2)}_{(\omega_{IR})}(z)$ between BIL and DL meanwhile indicates that the degree of preferential molecular orientation within these two regions must be vastly different.

More insight into the distribution of preferential water orientation with depth can be gained by analyzing the amplitude of the water signal as a function of z. Such a graph is depicted in Figure 5b, showing the depth-dependent peak maximum (at 3228 cm$^{-1}$) of the water band. From this graph it becomes evident that the preferential orientation slowly increases (exponentially) on approaching the interface and then undergoes a steep increase by two orders of magnitude at the transition from the DL to the BIL. This strong increase in preferential orientation seems, however, not sufficient to disturb the local H-bond structure (the H-bond strength and the orientational correlations of neighboring water molecules), as discussed above. This means that the extent of preferential molecular orientation even in the BIL must still be far from a perfectly aligned layer of dipoles because such a configuration would evidently induce significant distortion to the H-bond network. This conclusion is also clearly supported by the comparison of the amplitude of the water signal from the BIL to the signal from the C-H stretch vibrations. Despite the fact that there are roughly 10 times more water molecules in the BIL than $CH_3$ groups, we observe that the signals from latter are about twice as large as the water signal. This consequently suggests that there is a significant amount of cancellation between the signals from individual water molecules, signifying that their orientational distributions must be rather broad. This means that the picture of a well-ordered water layer underneath the surface charges with a saturated alignment of molecular dipoles (as previously suggested for such systems[67]) cannot be accurate for the investigated sample system. Instead, we must imagine the induced structural anisotropy as consisting of a macroscopic preference in molecular orientation where the ordering of water happens on a length-scale that is much larger than the orientational correlations in bulk water. In consequence, the deviations in thermodynamic properties of water in the interfacial region (compared to the bulk) are here clearly dominated by the lowering in entropy. Obviously, this somewhat surprising result cannot be generalized and is restricted to water in contact with the charged headgroups used in this study (phosphate) which are strongly hydrophilic. There are indications that the water structure shows more pronounced deviations in the BIL in other systems (e.g. more hydrophobic or electrode interfaces)[68–72], however, for a comprehensive analysis further investigations are required.

Overall, this analysis demonstrates the deep insight into the anisotropic interfacial water structure that can be obtained using the SFG/DFG technique. Importantly, such analysis can be done on any individual aqueous system without using spectral information from an additional measurement or the need for far-reaching assumptions and extended models. Although the presented mathematical description of the depth dependent response is closely adopted from the GCS theory, the presented results do not depend on the details of this model. There are, in fact, only two assumptions that are required to fully reconstruct the depth-dependent water response, namely i) the water anisotropy in the DL decays exponentially with the Debye length as the decay constant, and ii) the BIL is very thin (ca. 1nm) both of which are well justified. It should also be noted that even inaccuracies in the assumed thickness of the BIL would only moderately change the scaling in the obtained susceptibilities and thus the main conclusions presented here would not be affected. This renders the method little susceptible to model dependent bias which clearly sets it apart from other experimental approaches to address the depth-dependent nature of $\chi^{(2)}$ reported in recent publications.[22,24,26,45–47] Therefore, the presented technique provides a very promising perspective for future studies on charged aqueous interfaces, such as the investigation of the changes in the anisotropic water structure with different charges (density or different charged headgroups) at varying salt concentrations. More generally, further possible applications of the technique include depth-resolved studies at electrochemical interfaces, the investigation of ion-specific hydration effects, and the role of interfacial water in charge transfer processes. The obtained depth-resolution of the presented technique for such studies is particularly relevant as it allows for more precise investigation of the interplay between water structure and electrochemical properties, as well as providing insights into the dynamics of hydration and ion distribution in various systems.

## Conclusion

In this work we have presented a novel spectroscopic approach for the investigation of charged aqueous interfaces and demonstrated that it provides deep insight into the depth-dependent anisotropic water structure. The measurements of water in contact with charged (phosphate headgroup) surfactants at low salinity conditions ($10^{-5}$ M) reveal that the depth-dependent structural anisotropy of water is divided into two distinct regions, an induced preferential alignment of water dipoles in the bonded interfacial layer, followed by an exponentially decaying tail in the diffuse layer. At the transition between the two regions (from DL to BIL), the molecular alignment undergoes a sudden increase by two orders of magnitude. Despite this large difference in structural anisotropy, the obtained vibrational spectra show that the local hydrogen-bonding network remains largely unperturbed across the entire interfacial region in the investigated system, suggesting that the induced molecular alignment is even in the bonded interfacial layer small enough to not disturb the natural orientational correlations for neighboring molecules in bulk water. These findings establish a new paradigm for probing interfacial water, demonstrating that orientational ordering can coexist with intact hydrogen-bonding networks even under electrostatic influence. Our presented spectroscopic approach provides unprecedented depth sensitivity, paving the way for deeper insights into interfacial phenomena relevant to electrochemistry, biology, and environmental science.

## Acknowledgements

V.B. acknowledges financial support from the Comunidad de Madrid under the project "Cesar Nombela" 2023-T1/TEC-29119.

## Competing interests

There are no competing interests.

## Author contributions

A.P.F., V.B., and M.T. conceived the project and designed the experiments. A.D.D and A.P.F. performed the experiments and analyzed the data. All authors discussed the interpretation of the results. A.D.D and V.B. drafted the manuscript which was edited by all authors. M.W., and M.T. supervised the work and acquired funding.

## Data availability

Raw data will be made available upon reasonable request by contacting the corresponding author.

Anisotropic Water Structure at Charged Interfaces Studied by Depth-Resolved Vibrational SFG/DFG Spectroscopy

## Supplementary Information

Álvaro Diaz-Duque[1], Vasileios Balos[2], Martin Wolf[1], Alexander P. Fellows[1], and Martin Thämer[1*].

[1] Fritz-Haber-Institut der Max-Planck-Gesellschaft, Faradayweg 4-6, 14195, Berlin, Germany

[2] Instituto Madrileño de Estudios Avanzados en Nanociencia (IMDEA Nanociencia), 28049, Madrid, Spain

* Corresponding author

thaemer@fhi-berlin.mpg.de

(tel.): +49 (0)30 8413 5220

This supplementary information provides a detailed account of the experimental and analytical procedures employed to obtain the depth-resolved vibrational spectra discussed in the main text. First, a comparison of the different experimental approaches for phase-resolved SFG spectroscopy is shown and their suitability for depth-resolved studies is discussed. This is followed by additional details on the specific implementation of the time-domain interferometric setup used in this study. This includes additional information on the optical setup, the generation of local oscillators (LOs), and the strategies used to suppress parasitic background contributions, which are essential for accurate measurements at liquid interfaces. In a subsequent section, the mathematical framework underlying the time-domain reconstruction of vibrational SFG and DFG spectra is outlined. This includes the derivation of the delay-dependent signal and its transformation into spectrally resolved, phase-sensitive responses along the infrared axis. This is followed by information on the preparation of the charged surfactant monolayers at the air-electrolyte interface, including the control of surface coverage and the monitoring of monolayer formation. In the last part, a discussion on possible contributions from electric quadrupolar and magnetic dipole sources is presented along with additional information on the data analysis procedure. Together, these sections provide the theoretical and technical basis for the depth-resolved spectroscopic analysis of interfacial water presented in the main manuscript.

## A) Experimental

### A1) Comparison of different SFG approaches for SFG/DFG measurements

Established SFG spectrometers typically belong to one of the three following types: i) fully narrowband frequency-domain approaches,[1,2] here a narrowband IR beam is sequentially scanned over the frequency range of interest, ii) broadband IR frequency-domain approaches, where a broadband IR is mixed with a narrowband upconversion beam,[3,4] and iii) the recently introduced fully broadband time-domain method.[5,6] These experimental approaches to measure SFG spectra are relatively established from regular SFG measurements, however, they are not equally suitable for high precision phase-resolved SFG/DFG experiments on aqueous interfaces. Here we briefly discuss the theoretical and experimental aspects of these approaches in conjunction with such measurements.

In the frequency scanning approach, a narrowband IR beam is combined with an upconversion pulse to yield an SFG signal. This process is represented in the figure S-1a. The SFG signal has a certain spectral bandwidth that depends on the bandwidth of the visible (assuming that the IR frequency is a delta function) and an amplitude that scales with the nonlinear susceptibility at the given IR frequency. Obviously, in such a setting the spectrum of the generated nonlinear signal does not show any spectral features that relate to the vibrational resonances, instead this information is entirely encoded in its amplitude. By measuring the spectrally integrated SFG response on a single channel detector (SCD), while stepping through the infrared spectrum with the IR laser, the desired spectrum along the vibrational axis of the second-order susceptibility can be traced out (see figure S-1a). The frequency resolution obtained in the vibrational response here only depends on bandwidth of the IR beam (and obviously the step sizes), while the properties of the visible pulses are conceptually irrelevant. To obtain phase-resolved spectra with this approach, the interference of the nonlinear signal with a LO is analyzed. However, this process typically requires multiple measurements at different LO phases for each IR frequency.[7]

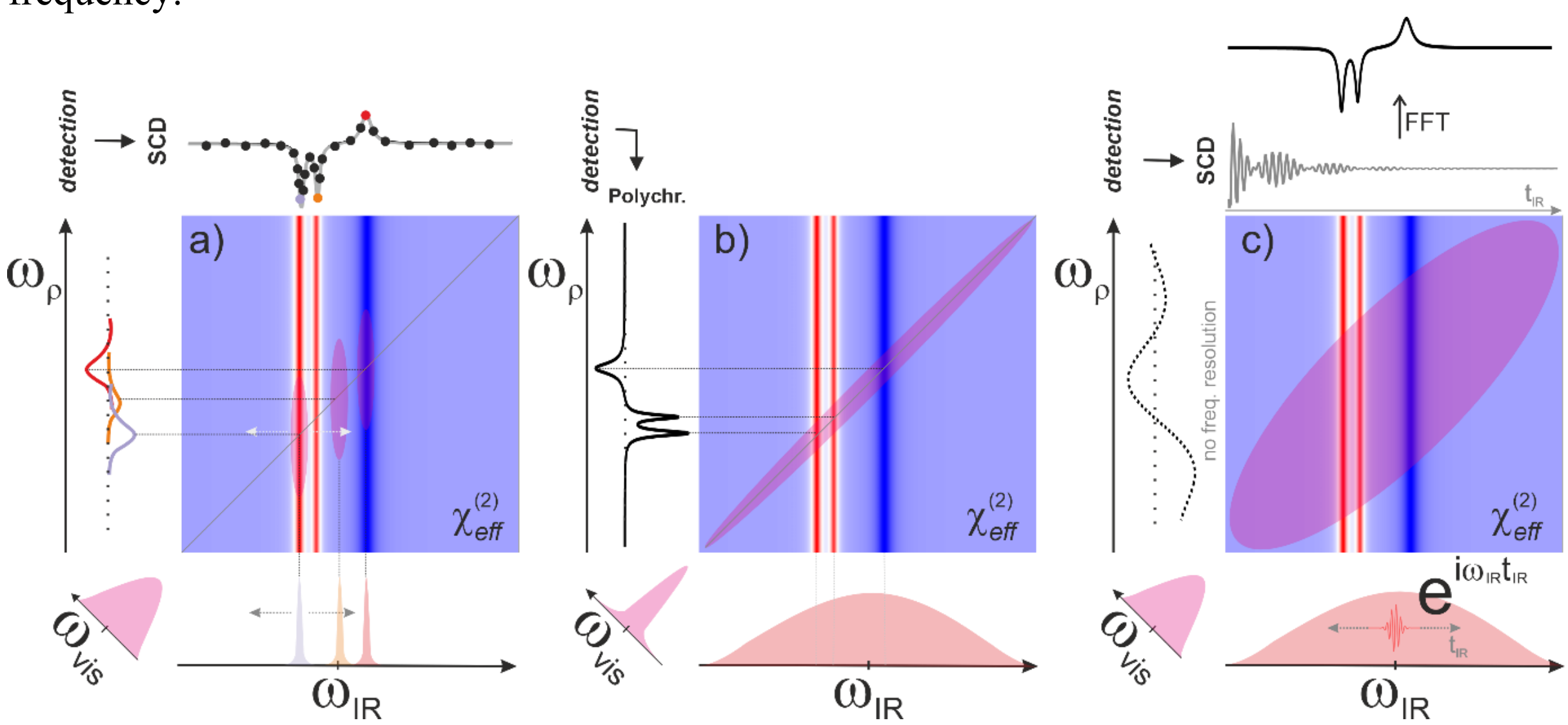

Figure S-1. Graphical representation of the three common approaches for performing frequency-resolved SFG measurements: a) frequency scanning approach, b) broadband IR frequency-domain approach, c) broadband time-domain method.

A widely used alternative approach uses broadband IR pulses in combination with narrowband visible pulses.[3,4] The interaction of the two beams with the sample is represented in figure S-1b. It can be seen that the spectrum of the generated SFG signal now directly contains the desired vibrational information of the sample. This is exploited in these methods by spectrally resolving the generated SFG light using a polychromator. The spectrum along the vibrational axis of $\chi^{(2)}$ can then be derived by downshifting the obtained frequency axis by the center frequency of the upconversion pulse. In contrast to the first SFG approach, the frequency resolution of the vibrational response is here limited by the bandwidth of the upconversion pulse. An ideal broadband frequency domain spectrometer therefore combines a maximum bandwidth IR pulse with a minimum bandwidth visible pulse. A significant advantage of this approach is the fact that phase-sensitive spectra can be obtained in a single measurement (except for need of a reference measurement for normalization). The LO is therefore delayed with respect to the SFG signal, giving rise to spectral interference fringes, which allow a precise determination of phase and amplitude of the resulting SFG spectrum.[3,4]

The third method is an interferometric time-domain approach employing broadband IR and broadband visible pulses (see figure S-1c), in combination with a LO.[5,8] The spectrum of the nonlinear signal does contain here the nonlinear vibrational information (all vibrational modes are excited by the broadband IR pulses), however, in contrast to the broadband frequency-domain approach, the spectral features are largely washed out by the bandwidth of the upconversion pulse. The desired frequency resolved spectrum along the vibrational axis of the two-dimensional susceptibility is here obtained using the concept of Fourier spectroscopy. By delaying the IR pulse with respect to the visible and LO pulses while detecting the intensity of the heterodyned nonlinear signal with a single channel detector, one obtains interference modulations between the generated SFG pulse and the LO. Fourier transformation of the resulting interferogram then yields the desired spectrum. The corresponding mathematical description of this time domain approach is shown in section A4.

Each method presented above can in principle be used for depth-resolved SFG/DFG measurements. However, the determination of the depth-dependent information requires enormous accuracy in phase, amplitude and frequency for both the SFG and the DFG responses, which cannot equally be achieved by the different approaches. As any deviation in one or more of these measured quantities from the correct sample responses will lead to large inaccuracies in the extracted depth profiles, it is essential that the experimental setup is adopted to these specific requirements.

The most challenging task among the requirements mentioned above is obtaining accurate and precise phase information for both the SFG and the DFG responses from a liquid sample that does not possess an atomically flat surface. The unavoidable presence of capillary waves[9] leads to constant fluctuations of the position of the phase boundary, which is also continuously influenced by the evaporation of the liquid. These spatial changes to the surface position introduce differences in the optical pathway, resulting in large phase variations of the generated nonlinear signals. An efficient way to avoid these phase fluctuations is to perform the experiment in a collinear geometry and with a LO that is generated before the sample.[6] That way the LO is linearly reflected by the same moving surface as the nonlinear signal, which leads to equal pathways for both and guarantees relative phase stability. Collinearity has the additional clear benefit that SFG and DFG signals are emitted in the same direction, which highly simplifies the interferometric measurement.

On the other hand, collinear geometries in nonlinear experiments have the clear drawback of the common appearance of parasitic signals. As all beams overlap on all optics inside the spectrometer, SFG and DFG signals can in principle be generated at all their surfaces. The use of spectral filters to remove these contributions is challenging because they would also remove the required LO. Traditionally this is overcome by avoiding spatial overlap of the beams on these optics, but this is obviously incompatible with collinearity. Another possibility is to avoid temporal pulse overlap inside the spectrometer, except at the sample surface. This can conveniently be done by introduction of dispersive material inside the beams just before the sample, which alters the delays between the infrared and upconversion pulses.[10] The additionally resulting temporal mismatch between the generated nonlinear signals and their respective LOs can subsequently be corrected after the sample using a birefringent crystal (see supplementary information for details). However, for this method to work in practice, visible and IR pulses must be short in time, which is evidently incompatible with narrowband pulses, as in the first two approaches. Only the interferometric time-domain uses exclusively short pulses and is therefore best suited for the implementation of a collinear beam geometry in combination with an efficient suppression of parasitic signals.

### A2) Generation of broadband infrared and vis pulses

The initial output from an amplified Ti:Sapphire laser (Astrella, Coherent), delivering pulses of 35fs duration at 800nm at a repetition rate of 1KHz, is split in two portions by a beam-splitter, before getting individually compressed in an internal and an external compressor. The two outputs pump two Optical Parametric Amplifiers (OPA; TOPAS Prime, Light Conversion), generating signal and idler pulses. For producing the visible output, the signal from the first OPA is isolated, and frequency doubled in a BBO crystal, yielding 690 nm light (visible). After removal of the fundamental using a short-pass filter (SPF), and subsequent power and polarization control using waveplates and polarizers, the beams are guided to the main interferometer. The mid-infrared (IR) pulses are generated by difference-frequency mixing between signal and idler outputs of the second OPA and sent to the interferometer. For adjustment of the time-overlap of the two pulses (IR and visible), the visible beam passes an automated delay line. Once time-overlap is found, the position of this delay-stage remains fixed during the whole experiment.

### A3) Generation of SFG and DFG local oscillators

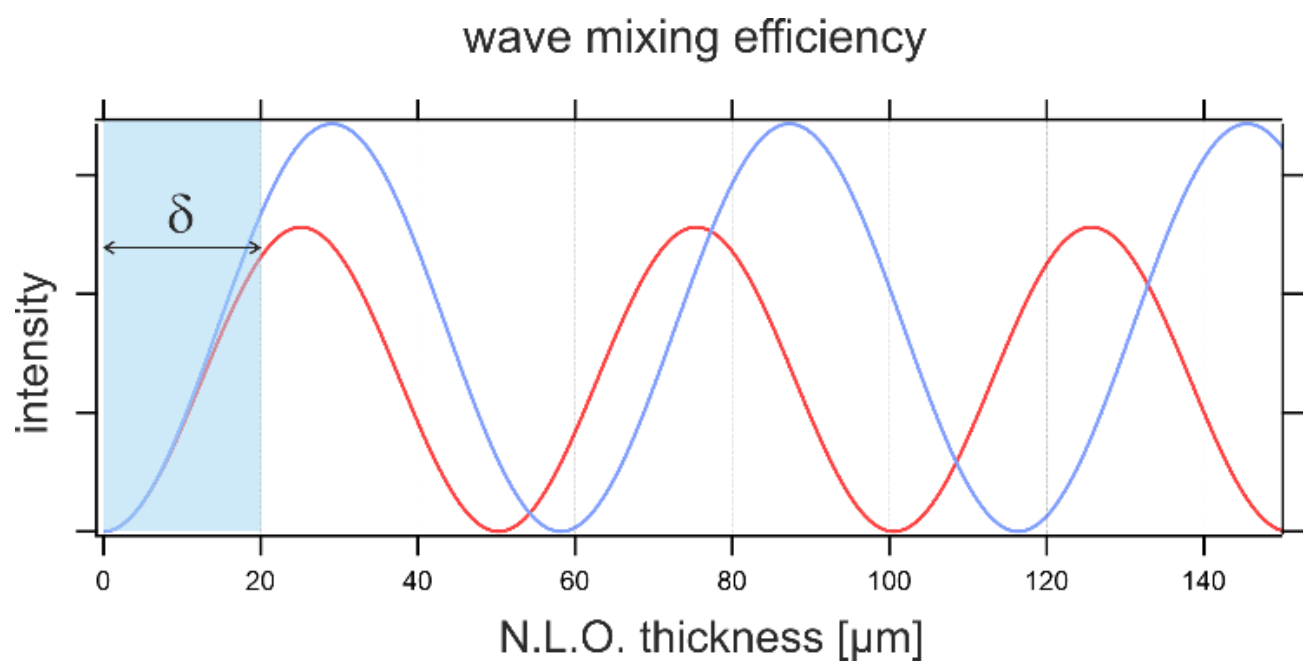

Figure S-2. Simulated SFG (red trace) and DFG (blue trace) local oscillator intensities as a function of z-cut quartz crystal thickness. The wavelengths used to calculate the curves are 690 nm for the visible beam and 3700 nm for the infrared beam.

The generation of the local oscillators (LO) for SFG and DFG is a cricial step for obtaining accurate phase resolved data with high signal-to-noise ratio. The intensity of the LO needs to generally match a relatively narrow intensity range for optimal performance of the balanced detection, which imposes important restrictions to the nonlinear material used for the LO generation. Choosing a z-cut alpha-quartz crystal has several advantages. Firstly, because of its crystal symmetry, the polarization of the LOs can precisely be tuned by simple azimuthal rotation around its z-axis (parallel to the beam direction), as long as the IR and visible pump beams are parallelly or perpendicularly polarized. Furthermore, the response of quartz is fully nonresonant and spectroscopically flat in the desired infrared region (between 1500 and 4000 $cm^{-1}$).[11] However, generating both LOs in transmition in quartz comes with restrictions on the crystal thickness (δ). Due to the absence of phase-matching, the generated nonlinear signals are subject to interference effects, which highly modulate the resulting intensities as function of crystal thickness, see equation S-2

Eq. S-1

$$I(\delta) = \delta^2 \mathrm{sinc}^2\left(\frac{\Delta k_z \delta}{2}\right)$$

where $\Delta k_z$ is the wavevector mismatch. This effect leads to oscillations of the generated LO intensity with crystal thickness (see figure S-2) with the oscillation period depending on the exact value for $\Delta k_z$. In principle it would be possible to maximize the LO intensity by choosing a crystal thickness at any maximum in this oscillating function, however, because of the dependency on $\Delta k_z$ the positions of the intensity maxima change as function of wavelength and, more importantly, bewteen SFG and DFG. This effect can create large intensity mismatches between the two LOs with highly negative impact on the signal-to-noise ratio. The only crystal thicknesses where the SFG and DFG LO intensities can be simultaneously maximized for all desired infrared frequencies (from 1500 and 4000 $cm^{-1}$) is in the range between approximatly 15 to 25 µm. For this reason, the z-cut quartz crystal used in our spectrometer has a thickness of 20 µm.

## A4) Elimination of parasitic signals

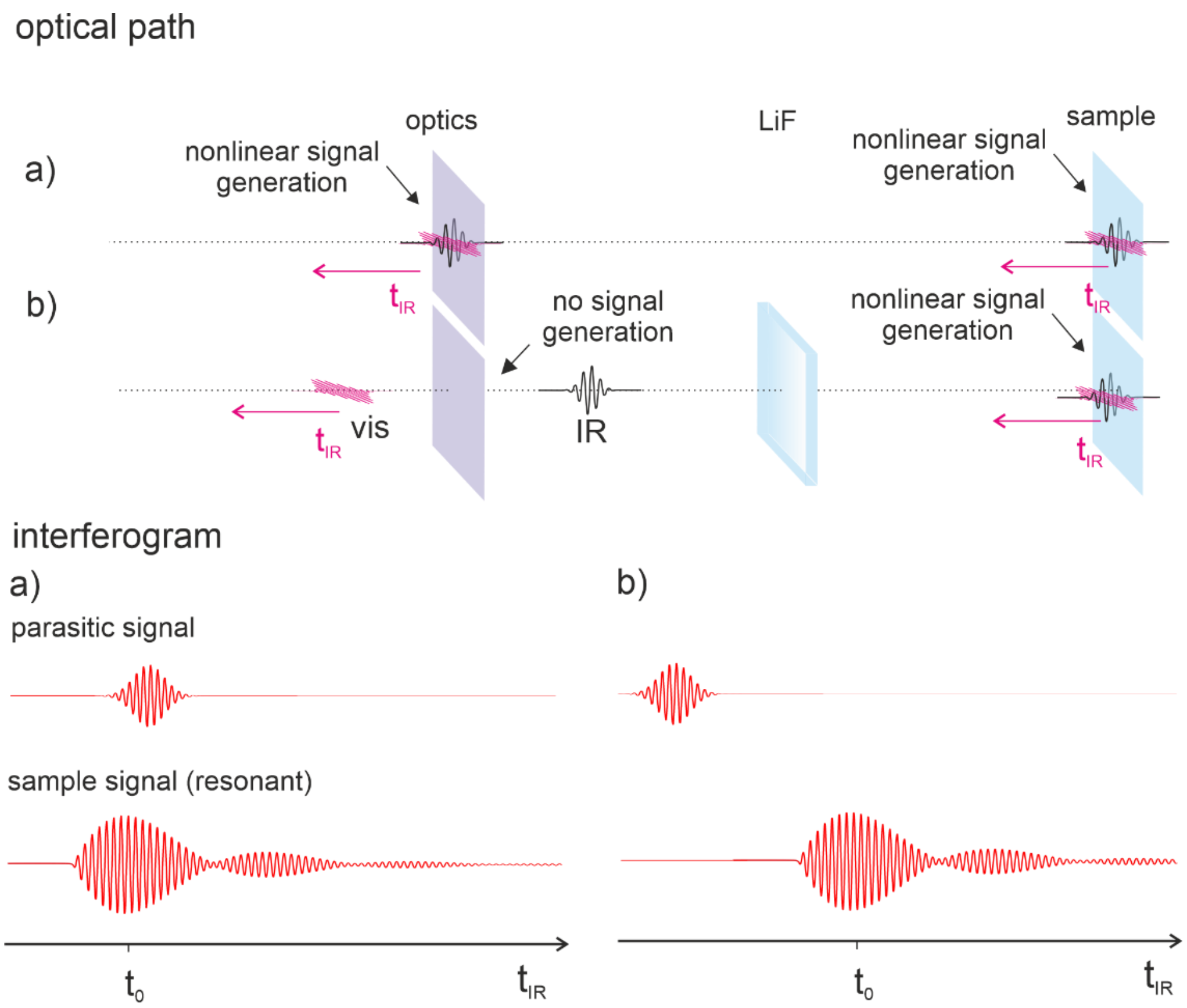

Figure S-3 Schematic representation of the optical paths (top) and resulting interferograms (bottom) a) without and b) with the introduction of a LiF window. In the latter case the interferogram arising from parasitic signal contributions generated by the optics along the beam path is shifted to negative delays by introducing a dispersive LiF window before the sample, exploiting its lower IR group velocity. This temporal displacement allows removing the parasitic signals from the measurement window.

An experimental drawback of our collinear geometry is that parasitic signals can be generated at several optics. Well-known sources of such parasitic signals are, for instance, the incoupling optics, as well as the focusing parabolic mirrors. Even though beams are only focused to generate the LO and the signal at the sample, weak parasitic signals can considerably affect the spectra from liquid interfaces. The reason for this is that aqueous interfaces normally generate extremely weak signals, and therefore the presence of parasitic signals can cause significant phase errors.

Parasitic contributions from the optics inside the interferometer generate interferograms just as does the sample response. However, these nonlinear responses from the optics are typically vibrationally nonresonant, making the resulting interferograms short (on the order of the IR autocorrelation). In contrast, the interferogram originating from the sample response can extend much further depending on the resonant spectrum, but only towards positive time delays. Therefore, the parasitic contributions can be effectively supressed by shifting the short parasitic interferograms to negative delay times, i.e. into a region outside the measured range. This can be achieved by, transmitting the beams just before the sample through a material that has a smaller group velocity at IR frequencies than at visible frequencies, such as LiF. The thickness of the LiF window can thereby be chosen to yield sufficient time delay for a specific infrared frequency. A schematic of this precedure is shown in figure S-3.

The temporal displacement of the different pulses inside LiF, however, also affects the relative timings of the two LOs with respect to the visible pulse. Due to the dispersion curve of LiF, the SFG LO now lags behind the visible pulse whereas the DFG LOs arrives earlier. This timing mismatch between the pulses highly reduces their interference efficiency between the LOs and the nonlinear signals generated at the sample interface which significantly reduces the amplitude of the interferogram. As shown in figure S-4 this challenge is overcome by the introduction of two calcite crystals in the detection path that exploit the orthogonal polarizations of the LOs with respect to the corresponding nonlinear signals to temporally overlap them.[10]

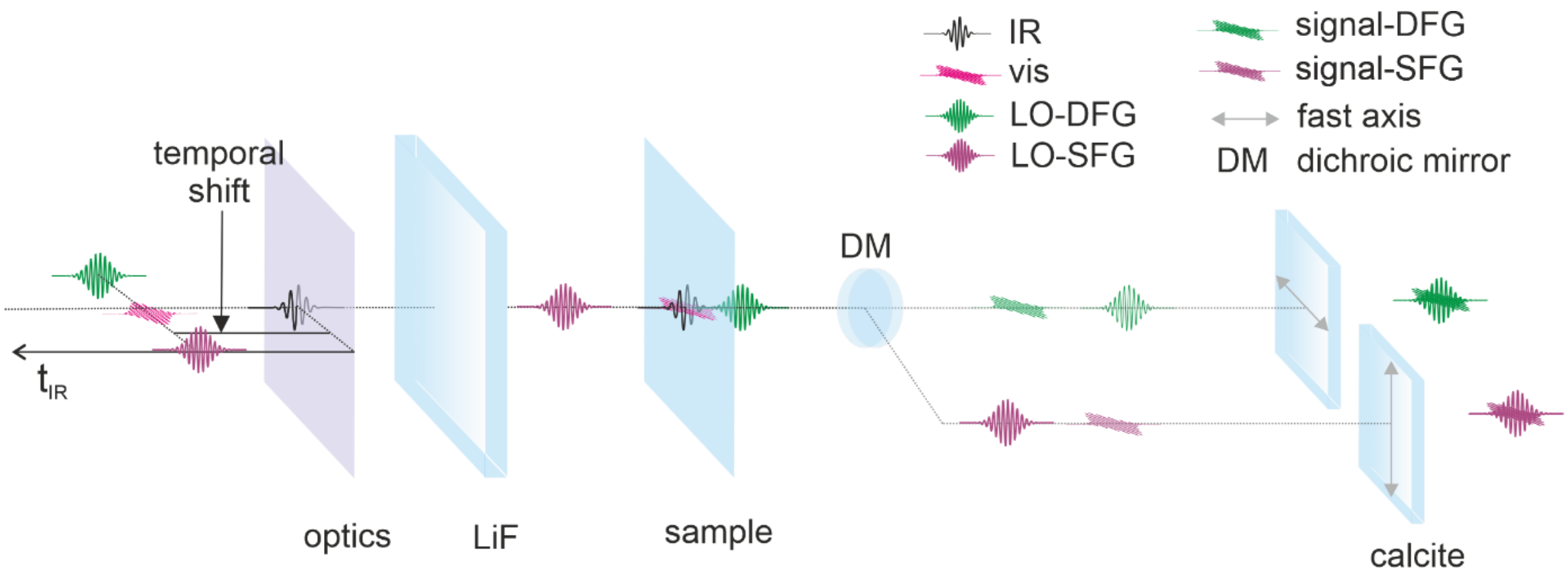

Figure S-4. Schematic showing the changes in relative timing between the different pulses while passing through the different optical ements intside the beam path. The dispersion in LiF introduces a mismatch between the visible pulse and the SFG/DFG local oscillators (LOs), reducing their interference efficiency between the LOs and the generated nonlinear signals. This mismatch is compensated downstream by a pair of birefringent calcite crystals that exploit the orthogonal polarizations of the LOs with respect to the nonlinear signals to temporally overlap them.

### A5) Time-domain acquisition of nonlinear vibrational spectra

As discussed in the main text, the nonlinear vibrational response of a non-centrosymmetric medium interacting with infrared (IR) and visible laser fields, denoted as $\widetilde{E}_{(\omega_{IR})}$ and $\widetilde{E}_{(\omega_{vis})}$, respectively, can be described by

Eq. S-2

$$\widetilde{E}_{(\omega_\rho)} \propto \int_{-\infty}^{\infty} d\omega_{IR} \int_{-\infty}^{\infty} d\omega_{vis} \widetilde{E}_{(\omega_{IR})} \widetilde{E}_{(\omega_{vis})} \chi^{(2)}_{(\omega_\rho=\omega_{IR}+\omega_{vis})} \cdot \delta_{(\omega_\rho-\omega_{IR}-\omega_{vis})}$$

where $\chi^{(2)}_{(\omega_\rho=\omega_{IR}+\omega_{vis})}$ is the effective second-order susceptibility that includes both the resonant molecular response and macroscopic propagation effects as well as nonlinear Fresnel factors.

In our time-domain approach, a controlled delay $t_{IR}$ is introduced between IR and visible pulses. This delay modulates the interaction of the visible beam with the vibrational free induction decay (FID) initiated by the IR excitation. The delayed interaction modifies the temporal evolution of the nonlinear polarization and the resulting expression for the radiated signal field becomes

Eq. S-3

$$\widetilde{E}_{(\omega_\rho,t_{IR})} \propto \int_{-\infty}^{\infty} d\omega_{IR} \int_{-\infty}^{\infty} d\omega_{vis} \widetilde{E}_{(\omega_{IR})} e^{i\omega_{IR}t_{IR}} \widetilde{E}_{(\omega_{vis})} \widetilde{E}_{(\omega_{vis})} \chi^{(2)}_{(\omega_\rho=\omega_{IR}+\omega_{vis})} \cdot \delta_{(\omega_\rho-\omega_{IR}-\omega_{vis})}$$

To access the full phase information of the nonlinear response, the signal is heterodyned with a local oscillator (LO) that is generated prior to the sample and co-propagates collinearly with the visible pulse. As a result, any temporal delay introduced between the IR and both visible pulses translates directly into a modulation of the phases between the nonlinear signal and the LO while their temporal envelopes are not affected. The resulting interference, recorded as a function of this delay, yields a time-domain interferogram that encodes both the amplitude and the phase of the effective susceptibility. This interference pattern can be expressed as

Eq. S-4

$$I_{(t_{IR})} = \int_{-\infty}^{\infty} dt\left(\mathbf{E}_{(t,t_{IR})} \cdot \mathbf{E}_{(t)}^{LO}\right) = \int_{-\infty}^{\infty} dt \int_{-\infty}^{\infty} d\omega_{\rho} \int_{-\infty}^{\infty} d\omega_{LO}\, \tilde{E}_{(\omega_{\rho},t_{IR})} \tilde{E}_{(\omega_{LO})}^{LO} e^{i(\omega_{\rho}+\omega_{LO})t}$$

where $\mathbf{E}_{(t,t_{IR})}$ and $\mathbf{E}_{(t)}^{LO}$ are the electric fields of the signal and LO in the time domain. The time integration in equation S-4 leads to a vanishing contribution unless the integrand contains matching frequency components. Consequently, only terms where $\omega_{\rho} = -\omega_{LO}$ contribute, simplifying the expression to

Eq. S-5

$$I_{(t_{IR})} = \int_{-\infty}^{\infty} d\omega_{\rho}\, \tilde{E}_{(\omega_{\rho},t_{IR})} \tilde{E}_{(\omega_{\rho})}^{*LO}$$

Substituting equation S-3 into equation S-5 yields

Eq. S-6

$$\mathbf{I}_{(t_{IR})} = \int_{-\infty}^{\infty} d\omega_{\rho} \int_{-\infty}^{\infty} d\omega_{IR} \int_{-\infty}^{\infty} d\omega_{vis} \tilde{E}_{(\omega_{\rho})}^{LO*} \tilde{E}_{(\omega_{IR})} e^{i\omega_{IR}t_{IR}} \tilde{E}_{(\omega_{vis})} \chi_{(\omega_{\rho}=\omega_{IR}+\omega_{vis})}^{(2)} \cdot \delta_{(\omega_{\rho}-\omega_{IR}-\omega_{vis})}$$

Fourier transformation of the interferogram then yields following expression:

Eq. S-7

$$S_{(\omega_S)} = \int_{-\infty}^{\infty} dt_{IR} \int_{-\infty}^{\infty} d\omega_{\rho} \int_{-\infty}^{\infty} d\omega_{IR} \int_{-\infty}^{\infty} d\omega_{vis} \tilde{E}_{(\omega_{\rho})}^{LO*} \tilde{E}_{(\omega_{IR})} e^{i(\omega_{IR}-\omega_S)t_{IR}} \tilde{E}_{(\omega_{vis})} \chi_{(\omega_{\rho}=\omega_{IR}+\omega_{vis})}^{(2)} \cdot \delta_{(\omega_{\rho}-\omega_{IR}-\omega_{vis})}$$

which can be written in a more compact form as:

Eq. S-8

$$S_{(\omega_{IR})} = \int_{-\infty}^{\infty} d\omega_{\rho} \int_{-\infty}^{\infty} d\omega_{vis} \tilde{E}_{(\omega_{\rho})}^{LO*} \tilde{E}_{(\omega_{vis})} \chi_{(\omega_{\rho}=\omega_{IR}+\omega_{vis})}^{(2)} \cdot \delta_{(\omega_{\rho}-\omega_{IR}-\omega_{vis})}$$

Here, $S_{(\omega_S)}$ was replaced by $S_{(\omega_{IR})}$ following the condition of a non-vanishing integral over $\omega_{IR}$ in equation S-7 imposing that $\omega_S = \omega_{IR}$. This step shows that the resulting spectrum is indeed a function of the vibrational frequency $\omega_{IR}$. Overall, the expression in equation S-8 is nothing else but the projection of the shaded area onto the vibrational frequency axis in figure 1 in the main text.

For samples, where the visible and emitted fields are non-resonant, their influence on the spectral shape of the second-order response is typically negligible.[12] As a result, the susceptibility can be approximated as a function that depends only on the IR frequency, scaled by a constant pre-factor. This simplification allows the susceptibility to be factored out of the frequency integrals and expressed as $\chi_{(\omega_{IR})}^{(2)}$. Under these conditions, equation S-8 becomes

Eq. S-9

$$S_{(\omega_{IR})} \propto \chi_{(\omega_{IR})}^{(2)} \tilde{E}_{(\omega_{IR})} \int_{-\infty}^{\infty} d\omega_{\rho} \int_{-\infty}^{\infty} d\omega_{vis} \tilde{E}_{(\omega_{\rho})}^{LO*} \tilde{E}_{(\omega_{vis})} \cdot \delta_{(\omega_{\rho}-\omega_{IR}-\omega_{vis})}$$

This final expression corresponds to the desired vibrational part of the susceptibility multiplied by an integral term that only depends on the complex spectra of visible and LO pulses. As the integral term is independent of $\chi^{(2)}_{(\omega_{IR})}$ it is eliminated by normalizing the sample response to a reference measurement and consequently does not affect the obtained spectra. The presence of this term, however, reveals another important aspect of the presented spectroscopic method. An alternative notation of equation S-9 is:

Eq. S-10

$$S_{(\omega_{IR})} \propto \chi^{(2)}_{(\omega_{IR})} \widetilde{E}_{(\omega_{IR})} \int_{-\infty}^{\infty} dt\, \widetilde{E}^{LO}_{(t)} \cdot \widetilde{E}^{vis}_{(t)} \cdot e^{i\omega_{IR}t}$$

which shows that the integral term is nothing else but the Fourier transform of the product of the visible and LO fields in the time domain evaluated at the frequency $\omega_{IR}$. This makes the amplitude of the obtained spectrum $S_{(\omega_{IR})}$ dependent on the temporal overlap between the visible and LO pulses. This dependency can be rationalized by the fact that the generated SFG pulse temporally coincides with the visible pulse which means that the above condition originates from the fact that SFG and LO pulses must temporally overlap. For best performance of the spectroscopic method in terms of signal-to-noise one should obviously maximize the amplitude of the interferogram which is why a precise matching of the arrival times of SFG and LO pulses is required. Within the presented experimental approach this is achieved by tuning the calcite crystals in the detection path presented in figure S-4.

Overall, these derivations provide the mathematical framework underlying the presented spectroscopic method. It shows how the time-domain approach, based on heterodyne detection and controlled delay between IR and visible pulses, yields phase-resolved spectra along the vibrational axis. This access to the complex spectral response forms the basis for retrieving depth-dependent structural information, as discussed in the main text.

## B) Sample preparation

The sample preparation details have been described elsewhere.[13,14] Briefly, the studied samples consist of insoluble surfactant monolayers deposited at the air–electrolyte interface. Electrolyte solutions were prepared by dissolving sodium chloride (NaCl, >99% purity, Sigma-Aldrich) in deionized water (Milli-Q) to yield concentrations of $10^{-1}$ and $10^{-5}$ M.

To form charged interfaces, insoluble lipids were employed. For positively charged interfaces, we used dihexadecyldimethylammonium bromide (DHAB, >97%, Sigma-Aldrich), while for negatively charged interfaces, dihexadecyl phosphate (DHP, Sigma-Aldrich ) was used. Both lipids were dissolved in chloroform to prepare spreading solutions with a concentration of 1 mg/mL. The solutions were stored at −20 °C until use.

Monolayers were prepared by dropwise deposition of the chloroform solutions onto the aqueous subphase using a micropipette, applying 2 µL per step until saturation was reached. For the presented studies it is important to work with a fully covered water surface since convection currents that are induced by heating with the infrared radiation (Marangoni convection) would otherwise drag the surfactants outside the illumination spot that is probed by our spectroscopy.[15] Saturation was identified by two simultaneous indicators. First, our fast-scan detection scheme (described in detail in the Experimental section) enables acquisition of SFG and DFG spectra within approximately 5 seconds. This allowed for real-time tracking of the appearance of the terminal methyl ($CH_3$) stretching vibrations around 2800 $cm^{-1}$. Second, at full surface coverage, an abrupt change in surface height occurs due to the lowering in surface tension. This change in apparent surface height was monitored using a height correction system based on a position-sensitive photodiode. This setup detects small deviations in beam alignment due to vertical movement of the liquid surface. The simultaneous appearance of $CH_3$ vibrational signatures and a discontinuity in surface height marks the completion of monolayer formation.

## C) Data treatment result section

### C1) Derivation of the mathematical expression for $\chi^{(2)}_{\mathrm{eff}}$

Starting from equation 6 in the main text

Eq. S-11

$$\chi^{(2)}_{eff,(\omega_\rho=\omega_{IR}+\omega_{vis})} = \int_0^\infty \mathrm{dz}\, \mathrm{L}_{(\omega_\mathrm{p})}\mathrm{L}_{(\omega_\mathrm{vis})}\mathrm{L}_{(\omega_\mathrm{IR})}\, \chi^{(2)}(\mathrm{z})\, \mathrm{e}^{\mathrm{i}\Delta \mathrm{k_z z}}$$

we assume that the nonlinear Fresnel factors $\mathrm{L}_{(\omega)}$ are independent of z which allows for placing the factors outside the integral. We then apply the non-resonant condition for the interactions at vis, SFG and DFG frequencies ($\chi^{(2)}(\mathrm{z}) \rightarrow \chi^{(2)}_{(\omega_\mathrm{IR})}(\mathrm{z})$) and obtain

Eq. S-12

$$\chi^{(2)}_{eff,(\omega_\rho=\omega_{IR}+\omega_{vis})} = \mathrm{L}_{(\omega_\mathrm{p})}\mathrm{L}_{(\omega_\mathrm{vis})}\mathrm{L}_{(\omega_\mathrm{IR})} \int_0^\infty \mathrm{dz}\ \chi^{(2)}_{(\omega_\mathrm{IR})}(\mathrm{z})\, \mathrm{e}^{\mathrm{i}\Delta \mathrm{k_z z}}$$

In our experiments we correct all the raw data for the nonlinear Fresnel factors using the two-layer model.[16] The effective susceptibilities $\chi^{(2)}_{\mathrm{eff}}$ presented in the result section therefore correspond to

Eq. S-13

$$\chi^{(2)}_{\mathrm{eff}} = \frac{\chi^{(2)}_{eff,(\omega_\rho=\omega_{IR}+\omega_{vis})}}{\mathrm{L}_{(\omega_\mathrm{p})}\mathrm{L}_{(\omega_\mathrm{vis})}\mathrm{L}_{(\omega_\mathrm{IR})}} = \int_0^\infty \mathrm{dz}\ \chi^{(2)}_{(\omega_\mathrm{IR})}(\mathrm{z})\, \mathrm{e}^{\mathrm{i}\Delta \mathrm{k_z z}}$$

To mathematically describe the evolution of the anisotropic response of the sample with depth we assume a 2-state model adopted from the GCS theory. This assumes the presence of two distinct structural regions. Firstly, a bonded interfacial layer BIL that includes the responses of the surfactants along with the first few water layers. The thickness of the water layer within this BIL is assumed to be approximately 1nm. As the functional form of the z-dependent response within this small thickness is unknown it is described by an effective average response $\chi^{(2)}_{\mathrm{BIL}}$. The second region (diffuse layer, DL) starting below this BIL is assumed to yield responses of constant spectral shape $\chi^{(2)}_{\mathrm{DL}}$ with an amplitude that decays exponentially with z. This assumption is reasonable because this region is sufficiently far from the phase boundary to exclude the possibility of i) hydrocarbons being present (surfactants are highly insoluble in water), and ii) significant changes in the water structure from solvation effects of the surfactant head groups. Furthermore, as the electric field in this region is relatively weak it can be expected that the water anisotropy (preferential molecular orientation) linearly follows the static electric field in that region. The nonlinear vibrational spectrum in this region should therefore resemble that of bulk water (with a small orientational anisotropy) and its amplitude should decay exponentially with the electrostatic field in z. The decay constant is meanwhile given by the Debye length following the prediction of the GCS model. With the above assumptions, we obtain the following expression for the evolution of the susceptibility with depth (Equation 7 in the main text)

Eq. S-14

$$\chi^{(2)}_{(\omega_{IR})}(z) = \begin{cases} \chi^{(2)}_{BIL} & 0 < z < z_{BIL} \\ \chi^{(2)}_{DL} \cdot e^{-(z-z_{BIL})/z_{DL}} & z \geq z_{BIL} \end{cases}$$

where $\chi^{(2)}_{DL}$ corresponds to the nonlinear response of the sample at the beginning of the DL (at $z = z_{BIL}$). Inserting equation S-14 into equation S-13 yields

Eq. S-15

$$\chi^{(2)}_{eff} = \int_0^{z_{BIL}} dz\ \chi^{(2)}_{BIL}\ e^{i\Delta k_z z} + \int_{z_{BIL}}^{\infty} dz\ \chi^{(2)}_{DL} \cdot e^{-(z-z_{BIL})/z_{DL}}\ e^{i\Delta k_z z}$$

Solving the integrals yields

Eq. S-16

$$\chi^{(2)}_{eff} = \chi^{(2)}_{BIL} \frac{1}{i\Delta k_z}\left(e^{i\Delta k_z z_{BIL}} - 1\right) + \chi^{(2)}_{DL} \frac{1}{1/z_{DL} - i\Delta k_z} e^{i\Delta k_z z_{BIL}}$$

Using the relation $|\Delta k_z z_{BIL}| \ll 1$ we find that

Eq. S-17

$$e^{i\Delta k_z z_{BIL}} \approx 1 + i\Delta k_z z_{BIL}$$

which yields

Eq. S-18

$$\chi^{(2)}_{eff} = \chi^{(2)}_{BIL}\, z_{BIL} + \chi^{(2)}_{DL}\, z_{DL} \cdot \left(\frac{1}{1 - i\Delta k_z z_{DL}} + \frac{i\Delta k_z z_{BIL}}{1 - i\Delta k_z z_{DL}}\right)$$

with $\Delta k_z z_{BIL} \ll 1$ we finally obtain

Eq. S-19

$$\chi^{(2)}_{eff} = \chi^{(2)}_{BIL}\, z_{BIL} + \chi^{(2)}_{DL}\, z_{DL} \cdot \frac{1}{1 - i\Delta k_z z_{DL}}$$

which can be transformed into

Eq. S-20

$$\chi_{\mathrm{eff}}^{(2)} = \chi_{\mathrm{BIL}}^{(2)} z_{\mathrm{BIL}} + \chi_{\mathrm{DL}}^{(2)} z_{\mathrm{DL}} \cdot \frac{1}{\sqrt{1+\Delta k_z{}^2 z_{\mathrm{DL}}{}^2}} e^{i\cdot \mathrm{atan}(\Delta k_z z_{\mathrm{DL}})}$$

This final equation equals equations 8 and 9 in the main text.

**C2) Quadrupolar and magnetic dipolar contributions**

The mathematical description of the nonlinear responses shown in the theory section of the main text only considers dipolar responses. However, beyond the electric dipole approximation additional electric quadrupolar and magnetic dipolar responses can in principle contribute to the measured signals. While the detailed theory of such responses in second order nonlinear spectroscopy can be found elsewhere in the literature[16,17] here we discuss possible quadrupolar and magnetic dipolar contributions to the measured spectra shown in this publication.

Electric quadrupolar and magnetic dipolar responses can roughly be divided into two categories: i) responses arising from the entire bulk which are induced by the derivative of the oscillating laser fields in the material (QB), and interfacial multipole responses arising from the derivative of the varying field amplitudes across the phase boundary (IQ). The contribution of the former is modulated by $\Delta k_z$ and will therefore give rise to different amplitudes in SFG and DFG responses but in contrast to electric dipolar sources from the bulk they do not induce any phase difference.[12] However, it has been shown that such bulk quadrupolar sources must vanish in SSP polarization measurements for collinear beam geometries used in this study.[16] In consequence there should be no QB contributions present in the recorded spectra.

The source of the IQ contribution is bound to the interface and therefore does not depend on $\Delta k_z$. This suggests that IQ responses may equally contribute to SFG and DFG spectra which would make them inseparable from the desired electric dipolar responses using the presented technique. However, very recently it was found that one portion of the overall IQ term, namely a magnetic dipolar contribution $\chi_{\mathrm{MD}}^{(2)}$ shows amplitude variations for SFG and DFG responses following the relations:[17]

Eq. S-21

$$\chi_{\mathrm{MD}}^{(2)\mathrm{SFG}} \sim \frac{\omega_{DFG}}{\omega_{SFG}} \; ; \; \chi_{\mathrm{MD}}^{(2)\mathrm{DFG}} \sim \frac{\omega_{SFG}}{\omega_{DFG}} \quad \rightarrow \quad \Delta\chi_{\mathrm{MD}}^{(2)} \sim \frac{\omega_{DFG}}{\omega_{SFG}} - \frac{\omega_{SFG}}{\omega_{DFG}}$$

In consequence such interfacial magnetic dipole signals contribute differently to the SFG and DFG spectra and based on these differences, $\chi_{\mathrm{MD}}^{(2)}$ can in principle be isolated and removed from the spectra. The residual IQ contributions in contrast remain experimentally inseparable.

Indications for the presence of such magnetic dipole contributions can be found in the spectra of the DHP sample at $10^{-1}$ M NaCl concentration shown in the main text in figure 4. The Debye length at this ion concentration is roughly 1 nm and DHP only decorates the phase boundary which should consequently lead to almost perfectly overlapping SFG and DFG spectra similar to what has been observed for the air-water interface presented in a recent publication.[12] Closer inspection of the SFG and DFG responses presented here shows that there are small but noticeable differences which strongly suggests that these differences indeed originate from the magnetic dipole contributions described above. In figure S-5 the difference spectrum for the DHP sample at $10^{-1}$ M NaCl concentration (red trace) is depicted. The spectrum shows that the deviations between SFG and DFG spectra mainly concern the C-H stretching region (mainly the pronounced peak at 2950 $cm^{-1}$) and must therefore originate from the DHP monolayer, more specifically from the hydrocarbon chains. In consequence, the same magnetic dipole response should appear in the measurement of the DHP sample at $10^{-5}$ M NaCl concentration. In figure S-5 the difference spectrum between SFG and DFG responses for latter measurement is shown (green trace) which confirms that the same spectral features in the CH region are indeed present along with the broad and phase shifted water response from the DL. With this data in hand the magnetic dipole contribution to the difference spectrum can be removed by simple subtraction of the $10^{-1}$ M from the $10^{-5}$ M trace as in the figure (black trace).

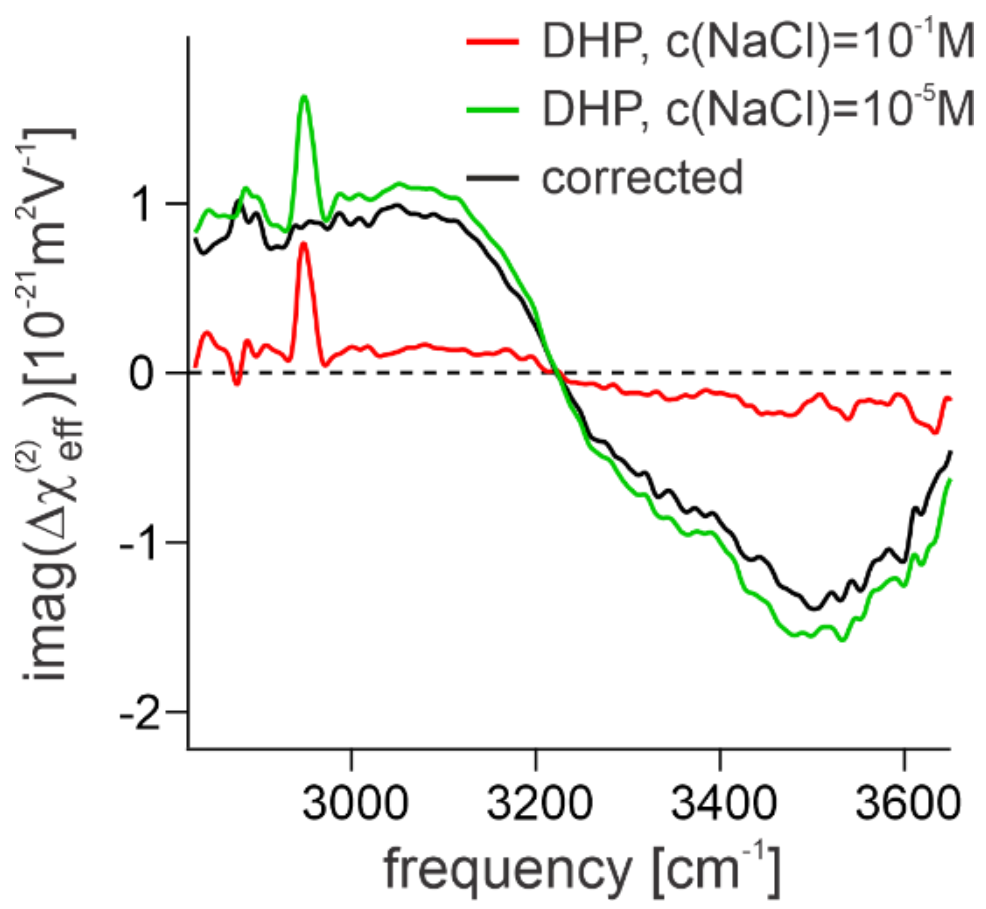

Figure S-5. SFG DFG difference spectra for DHP, c(NaCl) = $10^{-1}$M (red trace) and DHP, c(NaCl) = $10^{-5}$M (green trace). Black trace corresponds to the difference between the green and the red curves.

At this point it is important to note that the spectrum in figure S-5 (red trace) is the contribution of the magnetic dipole responses to the SFG/DFG difference spectra $\Delta\chi_{\mathrm{MD}}^{(2)}$ and not the contribution to the measured raw spectra ($\chi_{\mathrm{MD}}^{(2)\mathrm{SFG}}$ and $\chi_{\mathrm{MD}}^{(2)\mathrm{DFG}}$, respectively). Latter can be obtained based on equations S-21 via:

Eq. S-22

$$\chi_{\mathrm{MD}}^{(2)\mathrm{SFG}} = \frac{\Delta\chi_{\mathrm{MD}}^{(2)}}{1-\frac{\omega_{SFG}^2}{\omega_{DFG}^2}} \quad \text{and} \quad \chi_{\mathrm{MD}}^{(2)\mathrm{DFG}} = \frac{\Delta\chi_{\mathrm{MD}}^{(2)}}{\frac{\omega_{DFG}^2}{\omega_{SFG}^2}-1}$$

The resulting contributions for SFG and DFG are shown in figure 5 in the main text.

Besides the apparent magnetic dipole contribution additional signals from interfacial electric quadrupole sources can obviously contribute to the measured spectra. However, similar to the observed magnetic dipole signals they can be expected to be rather small and since they yield precisely equal SFG and DFG spectra they would exclusively contribute to the extracted BIL spectra.

## C2) Determination of DL and BIL spectra

For the determination of the DL spectrum we use the difference spectrum $\Delta\chi_{\mathrm{eff}}^{(2)}$ where the magnetic dipole contribution is removed (black trace figure S-5). Using equation 10 from the main text we can obtain the DL spectrum as follows

Eq. S-23

$$\chi_{\mathrm{DL}}^{(2)} \mathrm{z}_{\mathrm{DL}} = \frac{\Delta\chi_{\mathrm{eff}}^{(2)}}{\left(\mathrm{C}_{\mathrm{SFG}}(\mathrm{z}_{\mathrm{DL}}, \Delta\mathrm{k}_{\mathrm{z}}^{\mathrm{SFG}}) \cdot \mathrm{e}^{\mathrm{i}\cdot \mathrm{atan}(\Delta\mathrm{k}_{\mathrm{z}}^{\mathrm{SFG}}\cdot \mathrm{z}_{\mathrm{DL}})} - \mathrm{C}_{\mathrm{DFG}}(\mathrm{z}_{\mathrm{DL}}, \Delta\mathrm{k}_{\mathrm{z}}^{\mathrm{DFG}}) \cdot \mathrm{e}^{\mathrm{i}\cdot \mathrm{atan}(\Delta\mathrm{k}_{\mathrm{z}}^{\mathrm{DFG}}\cdot \mathrm{z}_{\mathrm{DL}})}\right)}$$

where $\chi_{\mathrm{DL}}^{(2)} \mathrm{z}_{\mathrm{DL}} = \int_{z_{BIL}}^{\infty} dz\, \chi_{\mathrm{DL}}^{(2)}$ (this spectrum is shown in figure 6a in the main text).

The effective contribution of the DL to the SFG and DFG spectra are obtained using following expression

Eq. S-24

$$\chi_{\mathrm{eff,DL}}^{(2)} = \chi_{\mathrm{DL}}^{(2)} \mathrm{z}_{\mathrm{DL}} \cdot \mathrm{C}_{\mathrm{SFG}}(\mathrm{z}_{\mathrm{DL}}, \Delta\mathrm{k}_{\mathrm{z}}) \cdot \mathrm{e}^{\mathrm{i}\cdot \mathrm{atan}(\Delta\mathrm{k}_{\mathrm{z}}\cdot \mathrm{z}_{\mathrm{DL}})}$$

with the respective values for $\Delta\mathrm{k}_{\mathrm{z}}$ for SFG and DFG, respectively. The resulting spectra are shown in figure 5 (main text).

The BIL contribution is then obtained by subtracting the effective DL contribution from the measured raw data (either from the SFG or the DFG data). Note, in contrast to the DL spectrum the BIL spectrum now still contains the magnetic dipole contribution, $\chi_{\mathrm{MD}}^{(2)}$ which is removed by simple subtraction. The final BIL spectrum ($\chi_{\mathrm{BIL}}^{(2)} z_{BIL} = \int_0^{z_{BIL}} dz\, \chi_{\mathrm{BIL}}^{(2)}$) is shown in figure 6a in the main text.